\documentclass[submission, Phys]{SciPost}

\usepackage[utf8]{inputenc} 
\usepackage[T1]{fontenc} 	
\usepackage[english]{babel} 

\usepackage[bitstream-charter]{mathdesign}
\usepackage{geometry} 		
\usepackage{amsmath} 		
\usepackage{mathtools} 		
\usepackage{subcaption} 
\usepackage{float} 			
\usepackage{graphicx} 		
\usepackage{tabularx} 		
\usepackage{booktabs} 		
\usepackage{color, xcolor} 	
\usepackage{pdfpages} 		
\usepackage{extarrows} 		
\usepackage{multirow} 		
\usepackage{multicol} 		
\usepackage{enumitem} 		
\usepackage{xspace} 		
\usepackage{stackrel} 		
\usepackage{tikz} 			
\usepackage{braket} 		
\usepackage{bm} 			
\usepackage{tensor} 		
\usepackage{slashed} 		
\usepackage{siunitx} 		
\usepackage{lastpage} 		
\usepackage{cite} 			
\usepackage[normalem]{ulem} 
\usepackage{fontawesome} 	
\usepackage{tocloft} 		
\usepackage{titlesec} 		
\usepackage{doi} 			
\usepackage{hyperref} 		
\usepackage[most]{tcolorbox} 					
\usepackage[nameinlink, capitalize]{cleveref} 	
\usepackage[nottoc, notlot, notlof]{tocbibind} 	
\usepackage[ruled, vlined]{algorithm2e} 		
\usepackage{makecell}
\usepackage[makeroom]{cancel}
\usepackage{feynmf}
\usepackage{dsfont}

\makeatletter
\def\BState{\State\hskip-\ALG@thistlm}
\makeatother

\makeatletter
\@ifundefined{pdfoutput}{}{\DeclareGraphicsRule{*}{mps}{*}{}}
\makeatother

\makeatletter
\DeclareRobustCommand*{\bfseries}{%
   \not@math@alphabet\bfseries\mathbf
   \fontseries\bfdefault\selectfont
   \boldmath
}
\makeatother

\hypersetup{
	pdftitle={Local Conformal Predictions for Calibrated Surrogates},
	pdfauthor={Dubey et al.},
	colorlinks=true, 			
	linkcolor={red!50!black}, 	
	citecolor={blue!50!black}, 	
	urlcolor={blue!80!black} 	
} 

\DeclareSymbolFont{usualmathcal}{OMS}{cmsy}{m}{n}
\DeclareSymbolFontAlphabet{\mathcal}{usualmathcal}

\SetArgSty{textnormal}
\SetKwComment{Comment}{{\small\#}~}{}
\SetCommentSty{mycommfont}

\setitemize{itemsep=0pt, parsep=0pt} 				
\setenumerate{itemsep=0pt, parsep=0pt} 				
\setitemize{itemsep=2pt,topsep=2pt,parsep=0pt,partopsep=0pt,leftmargin=*}
\setenumerate{itemsep=0pt,topsep=2pt,parsep=0pt,partopsep=0pt,labelindent=3pt,leftmargin=*}
\usepackage{amsmath}
 
\usepackage{amsthm} 		
\theoremstyle{definition}

\usepackage{placeins}
\usepackage{float}
\usepackage{graphicx}
\usepackage{pgffor}
 
\definecolor{red_cb}{HTML}{e41a1c}
\definecolor{blue_cb}{HTML}{377eb8}
\definecolor{green_cb}{HTML}{4daf4a}
\definecolor{purple_cb}{HTML}{984ea3}
\definecolor{orange_cb}{HTML}{ff7f00}

\definecolor{EmeraldGreen}{HTML}{1ea78d}
\definecolor{EnglishRed}{HTML}{b02427}
\hypersetup{colorlinks=true,urlcolor=EmeraldGreen,citecolor=EmeraldGreen,linkcolor=EnglishRed}

\newcommand{\ie}{\text{i.e.}\;}

\newcommand{\qqquad}{\qquad\quad}

\newcommand\one{\leavevmode\hbox{\small1\normalsize\kern-.33em1}}

\newcommand{\loss}{\mathcal{L}} 	

\newcommand{\madgraph}{\textsc{MadGraph5\_aMC@NLO}\xspace}

\newcommand{\arXiv}[2][]{%
	\ifthenelse{\equal{#1}{}}%
	{\href{http://arxiv.org/abs/#2}{arXiv:#2}}%
	{\href{http://arxiv.org/abs/#2}{arXiv:#2~[#1]}}}

\newcommand{\gev}{\text{GeV}}

\def\slashchar#1{\setbox0=\hbox{$#1$}           
   \dimen0=\wd0                                 
   \setbox1=\hbox{/} \dimen1=\wd1               
   \ifdim\dimen0>\dimen1                        
      \rlap{\hbox to \dimen0{\hfil/\hfil}}      
      #1                                        
   \else                                        
      \rlap{\hbox to \dimen1{\hfil$#1$\hfil}}   
      /                                         
   \fi}

\newcommand{\tikznode}[2]{%
\ifmmode%
\tikz[remember picture,baseline=(#1.base),inner sep=0pt] \node (#1) {$#2$};%
\else
\tikz[remember picture,baseline=(#1.base),inner sep=0pt] \node (#1) {#2};%
\fi}

\def\mathswitchr#1{\relax\ifmmode{\text{#1}}\else$\text{#1}$\xspace\fi}
\def\mathswitch#1{\relax\ifmmode#1\else$#1$\xspace\fi}

\graphicspath{{./figs/}}

\begin{document}

\begin{center}{\Large \textbf{
Local Conformal Predictions for Calibrated Surrogates}}
\end{center}

\begin{center}
  Suprio Dubey\textsuperscript{1,2}, 
  Henning Bahl\textsuperscript{1}, 
  Anja Butter\textsuperscript{1,3},
  J\"urgen Hesser\textsuperscript{2,4,5,6}, and
  Tilman Plehn\textsuperscript{1,4,6}
\end{center}

\begin{center}
{\bf 1} Institut f\"ur Theoretische Physik, Universit\"at Heidelberg, Germany \\
{\bf 2} Mannheim Institute for Intelligent Systems in Medicine, Universit\"at Heidelberg, Germany \\
{\bf 3} LPNHE, Sorbonne Universit\'e, Universit\'e Paris Cit\'e, CNRS/IN2P3, Paris, France \\
{\bf 4} Interdisciplinary Center for Scientific Computing (IWR), Universit\"at Heidelberg, Germany \\
{\bf 5} Central Institute for Computer Engineering (ZITI), Universit\"at Heidelberg, Germany \\
{\bf 6} CZS Heidelberg Initiative for Model-Based AI (MBAI), Universit\"at Heidelberg, Germany
\end{center}

\begin{center}
\today
\end{center}


\section*{Abstract}
     {\bf Neural network surrogates for LHC scattering amplitudes require trustworthy uncertainty estimates, a challenging task given the non-Gaussian systematics. We target it using conformal prediction, a distribution-free post-processing to complement trained surrogates with calibrated uncertainties. We find that standard conformal predictions struggle to provide locally calibrated uncertainties. This leads us to introduce FALCON, a novel conformal prediction method that learns locally calibrated confidence intervals. Our simple examples illustrate the power of distribution-free uncertainty quantification for ultra-fast event generation at the LHC.}

\vspace{10pt}
\noindent\rule{\textwidth}{1pt}
\tableofcontents\thispagestyle{fancy}
\noindent\rule{\textwidth}{1pt}
\vspace{10pt}

\clearpage
\section{Introduction}
\label{sec:intro}

The LHC has become the first precision hadron collider, where Run~3 and the High-Luminosity LHC will significantly improve the current experimental precision. Theory predictions must match this improvement. Here, modern machine learning (ML) is providing indispensable methods and tools~\cite{Butter:2022rso,Plehn:2022ftl} --- from accelerating phase-space sampling~\cite{Bothmann:2020ywa,Gao:2020zvv,Heimel:2022wyj,Bothmann:2023siu,Heimel:2023ngj,Deutschmann:2024lml,Heimel:2024wph,Janssen:2025zke,Bothmann:2025lwg,Heimel:2026hgp,Bothmann:2026dar} to generating full events~\cite{Hashemi:2019fkn,DiSipio:2019imz,Butter:2019cae,Butter:2021csz,Butter:2023fov,Brehmer:2024yqw,Butter:2024zbd,Favaro:2025pgz}.
These generative networks for LHC events are complemented by Monte Carlo simulations using network surrogates for the computationally expensive scattering amplitudes, the bottleneck of higher-order and high-multiplicity event generation~\cite{Bishara:2019iwh,Badger:2020uow,Aylett-Bullock:2021hmo,Maitre:2021uaa,Badger:2022hwf,Maitre:2023dqz,Brehmer:2024yqw,Bahl:2024gyt,Breso-Pla:2024pda,Favaro:2025pgz,Herrmann:2025nnz,Villadamigo:2025our,Beccatini:2025tpk,Bahl:2025xvx,Bahl:2026qaf,Bahl:2026jvt,DeCrescenzo:2026tsp}. 

The benefit of amplitude surrogates is that a trained network needs microseconds to provide a number for which the actual calculation takes minutes or even hours. However, a surrogate prediction without a reliable uncertainty estimate cannot be used for LHC physics. The two relevant questions are how accurately the network reproduces the true amplitude and how reliably the network tells us where it fails.

One simple answer is to train a Bayesian or heteroscedastic network that predicts a local uncertainty alongside the central amplitude~\cite{Badger:2022hwf,Bahl:2024gyt,Bahl:2025xvx,Bahl:2026qaf}. This method rests on the assumption that we know how to model the systematics and that the training data and the network expressivity can encode a calibrated uncertainty. Both of them are challenged exactly in the phase space regions where precision matters --- high-energy tails, soft and collinear regions, and the corners of phase space where event statistics are thin~\cite{Bahl:2026qaf}. The result is a familiar pathology: networks whose marginal coverage looks perfectly fine, while locally the truth escapes their error bar far more often than the nominal rate suggests.

A new way out is offered by conformal prediction~\cite{vovk2005, Shafer:2008tutorial, angelopoulos2023,Memmesheimer:2026abc}, only recently adopted in particle physics~\cite{Araz:2025vuw,Haussmann:2026gbi,Araz:2026img}. Here, we keep a calibration dataset, evaluate a measure of how badly the network missed the truth, and use the marginal quantiles of that measure to build a prediction interval. This interval comes with a finite-sample, distribution-free guarantee on the marginal coverage, independent of what the network is, what the residual distribution looks like, and how badly the predicted uncertainty was tuned. In this work, we explore and apply conformal prediction to surrogate amplitudes for LHC event generation. For particle physics, locally learned uncertainties are critical, motivating our new FALCON method to generalize global or marginal uncertainty quantification to local or conditional calibrated uncertainties.

In Sec.~\ref{sec:method} we set the stage and recall the distinction between global and local coverage. We introduce distribution-based and distribution-free uncertainty estimation, introduce conformal prediction, and present a series of established and novel implementations for global and local uncertainty estimation. In Sec.~\ref{sec:results_amp_regression}, we turn to a case study of gluon-associated $Z$ production applying various conformal prediction techniques. We summarize our findings in Sec.~\ref{sec:conclusions}.

\section{Calibrated uncertainties for surrogate amplitudes}
\label{sec:method}

An amplitude surrogate $A_\text{NN}(x)$ is trained to approximate a known $A_\text{true}$ in every phase-space point $x$,
\begin{align}
    A_\text{NN}(x) \approx A_\text{true}(x) \; .
\end{align}
This approximation is subject to uncertainties: systematic or irreducible uncertainties plateau towards infinite amounts of training data; statistical or reducible uncertainties decrease with more training data. In this study we focus on systematics.

\subsection{Confidence intervals}

To estimate the systematic uncertainties of $A_\text{NN}(x)$, we construct confidence intervals $C_\alpha(x)$ such that with a probability of $(1-\alpha)$ the true amplitude lies within $C_\alpha(x)$.

We test its calibration using two coverage metrics:
\begin{enumerate}
\item \textbf{Global or marginal coverage} is the probability that $A_\text{true}(x)$ lies in $C_\alpha(x)$, integrated over the whole phase space $x$,
\begin{align}
    c^\text{marg}_{1-\alpha} = \int dx\, p(x)\, P(A_\text{true}(x) \in C_\alpha(x)) \; ,
    \label{eq:global_coverage}
\end{align}
where $p(x)$ encodes the phase space probability. We estimate the marginal coverage for $N$ phase space points as 
\begin{align}
  c_{1-\alpha}^\text{marg} 
  = \frac{1}{N}
  \sum_{i = 1}^{N}
  \mathds{1}\!\left\{A_\text{true}(x_i) \in C_\alpha(x_i)\right\},
  \label{eq:marginal_coverage}
\end{align}
where $\mathds{1}\{\cdot\}$ is the indicator function --- $1$ if the condition holds or $0$ otherwise. Calibrated global confidence intervals require
\begin{align}
 c^\text{marg}_{1-\alpha} = 1 - \alpha
 \qqquad \text{for every $\alpha$.}
\end{align}
The marginal coverage will detect a global mis-calibration of the confidence intervals, but no local failures. For instance, the confidence interval may be too wide in simple-to-fit regions and too narrow in hard-to-fit regions, such that both failure modes cancel. 

\item \textbf{Local or conditional coverage} evaluates the coverage locally in $x$ and should be
\begin{align}
    c^\text{cond}_{1-\alpha}(x) = P\!\left(A_\text{true}(x) \in C_\alpha(x) \right)
    = 1 - \alpha\; ,
    \label{eq:local_coverage}
\end{align}
now for every $\alpha$ and for every $x$. To compute it, we bin the phase space and evaluate 
\begin{align}
  c^\text{cond}_{1-\alpha}(x) =  
  P\!\left(A_\text{true}(x) \in C_\alpha(x)
                    \mid x \in \text{bin} \right) \; .
  \label{eq:bin_local_coverage}
\end{align}
Finer binning converges to the continuum result with increasing statistical fluctuations.
\end{enumerate}
The two coverages are related via
\begin{align}
  c^\text{marg}_{1-\alpha}
  = \int dx \; p(x) \; c^\text{cond}_{1-\alpha}(x) \; .
\end{align}
For LHC physics, the local coverage is crucial, as phase-space regions with challenging kinematics can have poor coverage that the marginal coverage misses entirely.

\subsubsection*{Gaussian likelihood}
\label{sec:Gaussian_baseline}

To construct confidence intervals, we can phrase the surrogate training as minimization of the negative log-likelihood and assume a Gaussian likelihood in terms of the local mean $A_\text{NN}(x)$ and the local systematic variance $\sigma^2(x) \equiv \sigma^2_\text{syst}(x)$~\cite{Plehn:2022ftl}. This ansatz works well for amplitudes~\cite{Bahl:2024gyt,Bahl:2025xvx} and leads to the heteroscedastic loss
\begin{align}
    \loss_\text{het} = 
        \frac{[A_\text{NN}(x) - A_\text{true}(x)]^2}{2\sigma^2(x)} + \log \sigma(x) \; .
    \label{eq:het_loss}
\end{align}
At each phase space point, the training minimizes the numerator, but the network can also increase $\sigma(x)$ at the price of the logarithmic term. The Gaussian likelihood allows us to compute the two-sided $p$-value
\begin{align}
    p \bigl(A_\text{true}(x) \,\big|\, A_\text{NN}(x),\,\sigma(x)\bigr)
    \;=\; 2\!\left[\,1 - \Phi\!\left(\frac{|A_\text{NN}(x) - A_\text{true}(x) |}{\sigma(x)}\right)\right] \; ,
    \label{eq:gaussian_pvalue}
\end{align}
where $\Phi$ is the standard normal cumulative distribution function. The confidence interval at nominal level ($1-\alpha$) collects all amplitude values whose $p$-value exceeds $\alpha$,
\begin{align}
    C_\alpha (x)
    &= \bigl[A_\text{NN}(x) - \Phi^{-1}(1-\alpha/2) \,\sigma(x),\;
             A_\text{NN}(x) + \Phi^{-1}(1-\alpha/2) \,\sigma(x)\bigr] \; .
    \label{eq:gaussian_interval}
\end{align}
The calibration of these confidence intervals rests on the Gaussian assumption. If the true residual distribution is skewed or heavy-tailed, we can model the systematics using a Gaussian mixture~\cite{Bahl:2024gyt,Bahl:2025xvx,Bahl:2026qaf,ATLAS:2024rpl} or a Student's $t$-likelihood~\cite{Bahl:2026qaf}.

\subsection{Quantile regression}
\label{sec:quantile_regression}

Without assuming a likelihood, we can directly target the confidence intervals through quantile regression~\cite{Koenker:1978reg}. Rather than a central value $A_\text{NN}(x)$ and a standard deviation $\sigma(x)$, we learn functions $q_\tau(x)$ for quantile levels $\tau \in (0,1)$, such that a proportion $\tau$ of the true amplitudes lies below $q_\tau(x)$,
\begin{align}
    P(A_\text{true}(x) < q_\tau(x)) = \tau\;.
\end{align}
A network is trained on a series of quantile levels $\{\tau_1, \ldots, \tau_Q\}$ to encode a general posterior in terms of local prediction intervals $[q_{\tau_{\text{lo}}}(x), q_{\tau_{\text{hi}}}(x)]$.

\subsubsection*{Pinball loss}

To learn the local quantile $q_\tau$ for a fixed level $\tau \in (0,1)$ we can use the pinball loss 
\begin{align}
    \loss_\tau = 
    \begin{cases}
      \tau\, \left[A_\text{true}(x) - q_\tau(x) \right]
      & \text{if } A_\text{true}(x) \geq q_\tau \\
      (1 - \tau) \,\left[ q_\tau(x)  - A_\text{true}(x) \right]
      & \text{if } A_\text{true}(x) < q_\tau \; .
    \end{cases}
    \label{eq:pinball}
\end{align}
For $\tau > 0.5$, underestimates $A_\text{true}(x) \geq q_\tau$ are penalized more than overestimates, so $q_\tau$ is pushed up until the proportion of training targets falls below $\tau$. For several quantile levels $\{\tau_1, \dots, \tau_Q\}$ the loss is the combination
\begin{align}
    \loss_\text{pinball} = \frac{1}{Q}\sum_{i=1}^{Q} \loss_{\tau_i}(x) \; .
    \label{eq:pinball_total}
\end{align}
The quantile network predicts an interval for the nominal value $1-\alpha$ as
\begin{align}
  C_\alpha(x)
  = \bigl[q_{\tau_{\text{lo}}}(x),\, q_{\tau_{\text{hi}}}(x)\bigr]
  \qquad \text{with} \qquad
  \tau_{\text{lo}} = \frac{\alpha}{2}
  \qquad
  \tau_{\text{hi}} = 1 - \frac{\alpha}{2} \; .
  \label{eq:qr_interval}
\end{align}
We can evaluate the coverage in Eq.\eqref{eq:marginal_coverage}, and if the learned quantiles are correct, we expect
\begin{align}
  P(A_\text{true}(x) < q_{\tau_{\text{lo}}}(x))
  = P(A_\text{true}(x) > q_{\tau_{\text{hi}}}(x)) 
  &= \frac{\alpha}{2} \notag \\
  \Rightarrow \qqquad 
  c^\text{cond}_{1-\alpha}(x) 
  &= 1 - \alpha \; .
\end{align}
Because each quantile is a function of $x$, the width $q_{\tau_{\text{hi}}}(x) - q_{\tau_{\text{lo}}}(x)$ will vary over phase space. Moreover, the quantile distances from the median can differ,
\begin{align}
    q_{\tau_{\text{hi}}}(x) - q_{0.5}(x) \neq q_{0.5}(x) - q_{\tau_{\text{lo}}}(x)\; .
\end{align}
%

\subsection{Naive conformal prediction}
\label{sec:conformal}

Conformal prediction (CP) calibrates confidence intervals of a trained surrogate using an additional dataset. It starts from a function that measures how badly the surrogate misses the truth. The simplest choice is the residual
\begin{align}
  R(x) = \bigl|A_\text{true}(x) - A_\text{NN}(x)\bigr|\;.
  \label{eq:test_residual}
\end{align}
For a symmetric confidence interval,
\begin{align}
    C_\alpha(x) = \bigl[A_\text{NN}(x) - q_\alpha,\; A_\text{NN}(x) + q_\alpha\bigr]\;,
    \label{eq:cp_interval}
\end{align}
we can write the global calibration requirement of Eq.\eqref{eq:global_coverage} as
\begin{align}
  \int dx\,p(x)\,P(R(x) \leq q_\alpha) = 1 - \alpha \; .
  \label{eq:cp_goal}
\end{align}

We extract $q_\alpha$ from a calibration dataset of the same form as the surrogate training data. We order its data points by ascending residual,
\begin{align}
    R_{\text{calib},1} < \ldots < R_{\text{calib},N}
    \qquad \text{with} \qquad 
    R_{\text{calib},i} = R(x_{\text{calib},i})\; .
\end{align}
A new data point with $R_\text{test} = R(x_\text{test})$ sampled from the same distribution is exchangeable with the calibration data points and is equally likely to take any of the combined $(N+1)$ positions ordered by residual,
\begin{align}
    P\left(R_\text{test} \leq R_{\text{calib},k}\right)
    &= \frac{k}{N + 1}
    \qquad \text{for any} \quad  k = 1,\ldots, N+1 \; .
    \label{eq:coverage_derivation2}
\end{align}
For given $(1-\alpha)$ we pick the smallest integer $k_\alpha$ for which the left-hand side reaches this value,
\begin{align}
    k_\alpha = \text{ceiling}\left[(1 - \alpha)(N+1)\right]\;,
\end{align}
The `ceiling' operation rounds to the next integer, $a \le \text{ceiling}\left[ a \right] < a + 1$. This yields the two-sided bound
\begin{align}
    1 - \alpha \;\le\; P\left(R_\text{test} \leq R_{\text{calib},k_\alpha}\right) < 1-\alpha + \frac{1}{N+1}
    \label{eq:cp_sandwich}
\end{align}
and leads us to identify
\begin{align}
    q_\alpha = R_{\text{calib},k_\alpha}\; .
    \label{eq:qa_def}
\end{align}
The lower bound in Eq.\eqref{eq:cp_sandwich} becomes the so-called coverage guarantee of Eq.\eqref{eq:cp_goal}, and the upper bound implies that the interval over-covers by at most $1/(N+1)$. When $k_\alpha > N$, so the calibration dataset is too small to supply the requested quantile, the residual $R_{\text{calib},k_\alpha}$ does not exist, and we set $q_\alpha = +\infty$.

This construction of the confidence intervals also works for other functions than the absolute residual in Eq.\eqref{eq:test_residual}, for example, the signed residual
\begin{align}
 R_s(x) = A_\text{true}(x) - A_\text{NN}(x) \; ,
\end{align}
which results in the one-sided confidence interval $(-\infty,\, A_\text{NN}(x) + q_\alpha]$.

\subsection{Adaptive conformal prediction}
\label{sec:adaptive}

Starting from Eq.\eqref{eq:cp_sandwich}, we are free to design orderings that incorporate additional, local information. We use two such constructions, ordered by increasing adaptivity: $(i)$ a scaled or heteroscedastic residual (Het) and $(ii)$ conformalized quantile regression (CQR). Despite their local nature, the uncertainty bands derived in this way are not guaranteed to be calibrated locally.

\subsubsection*{Heteroscedastic residual (Het)}

Our first adaptive construction includes a local uncertainty estimate, producing intervals whose width varies across phase space according to the local uncertainty~\cite{article, papadopoulos2011}. Based on the heteroscedastic loss from Sec.~\ref{sec:Gaussian_baseline}, the nonconformity measure
\begin{align}
    R_\sigma(x) = \frac{|A_\text{true}(x) - A_\text{NN}(x)|}{\sigma(x)}
    \label{eq:scaled_residual_score}
\end{align}
gives the prediction interval
\begin{align}
    C_\alpha(x) =
    \left[A_\text{NN}(x) - q_\alpha\,\sigma(x), A_\text{NN}(x) + q_\alpha\,\sigma(x)\right] \; .
    \label{eq:prediction_set_scaled}
\end{align}
Its width varies with $x$ in proportion to the heteroscedastic uncertainty, but the confidence intervals remain symmetric around the central prediction.

The Gaussian assumption used to define $\sigma(x)$ does not affect the calibration. If the likelihood is non-Gaussian, $\sigma(x)$ is mis-specified, and the intervals are less tightly adapted to the local distribution. This reduces the local coverage but not the global one. The conformal calibration step constructs $q_\alpha$ to absorb any mis-calibration in $\sigma$, so the marginal coverage at level $1-\alpha$ holds regardless of the true residual distribution. This nonconformity measure inherits the local adaptivity of the Gaussian confidence intervals while restoring the distribution-free coverage guarantee.

\subsubsection*{Conformalized quantile regression (CQR)}

Conformalized quantile regression~\cite{NEURIPS2019_5103c358} removes the restriction of symmetric confidence intervals. Following Sec.~\ref{sec:quantile_regression}, we train a single quantile regression network to encode multiple local quantiles $\{\tau_1, \ldots, \tau_Q\}$. A pair $q_{\tau_{\text{lo}}}(x)$ and $q_{\tau_{\text{hi}}}(x)$ define the lower and upper edges of an initial confidence interval, which we use to define a nonconformity measure
\begin{align}
    R_\text{CQR}(x) = \max\!\left[ q_{\tau_{\text{lo}}}(x) - A_\text{true}(x),\;
    A_\text{true}(x) - q_{\tau_{\text{hi}}}(x)\right] \; .
    \label{eq:cqr_score}
\end{align}
It is positive when $A_\text{true}(x)$ falls outside the predicted confidence interval and negative when it is within. Conformal calibration then gives us the confidence interval
\begin{align}
    C_\alpha(x) = \left[q_{\tau_{\text{lo}}}(x)
    - q_\alpha,\; q_{\tau_{\text{hi}}}(x) +
    q_\alpha\right] \; ,
    \label{eq:cqr_interval}
\end{align}
where the quantile regression interval is expanded by the constant calibration parameter $q_\alpha$ and inherits its shape from the underlying local quantile band. It is asymmetric whenever the amplitude distribution is skewed. The conformal step only adds the marginal coverage guarantee, valid without any assumptions about the residual distribution.

Even with these two adaptive nonconformity measures, the coverage 
\begin{align}
    P(A_\text{true}(x) \in C_\alpha(x)) = 1 -\alpha 
\end{align}
remains global. The calibration parameter $q_\alpha$ is fixed by the bulk of the calibration distribution, where the surrogate is typically accurate. 

\subsection{Local calibration}

Local uncertainty bands are crucial for LHC physics. Critical kinematic regions like high-energy tails or threshold structures contribute only a small fraction of the calibration sample. This requires proper local conformal prediction and leads us to develop a new, field-adapted local conformal prediction (FALCON).

\subsubsection*{Mondrian conformal prediction}

One way to achieve at least partial local coverage is Mondrian conformal prediction~\cite{fe5374fa2e6b4c269718c6de868bab26, Vovk:2012conditional}. We partition the phase space into fixed volumes $V_i$ with $N_i$ calibration points and $\{R^{(i)}_{\text{calib},j}\}_{j=1}^{N_i}$ as the nonconformity measure. The partitioned calibration gives us 
\begin{align}
  q_\alpha^{(i)} = R^{(i)}_{\text{calib},k_\alpha^{(i)}} \; .
  \label{eq:mondrian_qhat}
\end{align}
The per-volume conformal prediction algorithm then satisfies
\begin{align}
  P\!\bigl(A_\text{true}(x) \in C_\alpha^{(i)}(x) \mid x \in V_i\bigr)
  = 1 - \alpha \; .
  \label{eq:mondrian_guarantee}
\end{align}
The price for the local uncertainty estimate is a massively reduced statistic. Typical failure regions are sparsely covered, with only a few calibration points falling there, and the per-cell quantile estimate becomes noisy. Targeted sampling~\cite{Bahl:2026qaf} can resolve this issue to some extent, but does not change the scaling problem of the partitioning. If Mondrian partitioning bins hold too few points to calibrate well, we need another route.

\subsubsection*{Field-adapted local conformal prediction (FALCON)}
\label{sec:falcon}

To implement a local correction across phase space with an improved scaling with dimensionality, we define probe regions $\mathcal{H}_i$, compute $q_\alpha^{(i)}$ in each of them, and interpolate
\begin{align}
 \tilde{q}_\text{inter}(x) = \text{ linear interpolation of } q_\alpha^{(i)}\;.
\end{align}
If the surrogate and the true amplitude are smooth, $\tilde{q}_\text{inter}(x)$ approximately follows the correction everywhere, including sparse tails. 

A problem with this new construction is that the marginal coverage guarantee does not necessarily hold. Simply replacing $q_\alpha$ by $\tilde{q}_\text{inter}$ violates the guarantee, as $q_\alpha$ needs to be a fixed scalar, not a function of $x$. To preserve the guarantee while using $\tilde{q}_\text{inter}$, we integrate $\tilde{q}_\text{inter}$ into the nonconformity measure instead of simply replacing $q_\alpha$ with $\tilde{q}_\text{inter}$. Based on the above discussion, we define two FALCON variations:
\begin{enumerate}
\item We first discuss the case of building the interpolation on top of the heteroscedastic residual function. To adapt the procedure to the form of Eq.\eqref{eq:scaled_residual_score}, which normalizes by $\sigma$, we divide by $\tilde q_\text{inter}$,
\begin{align}
  R'_\sigma(x) = \frac{R_\sigma(x)}{\tilde{q}_\text{inter}(x)}\;.
  \label{eq:falcon_renorm_scaled}
\end{align}
Using $R'_\sigma$ to derive $q_\alpha$ restores the marginal coverage guarantee for the interval
\begin{align}
    C_\alpha^{\text{FALCON-het}}(x)
    = \bigl[\,A_\text{NN} - q_\alpha\,\tilde{q}_\text{inter}\,\sigma,\;
          A_\text{NN} + q_\alpha\,\tilde{q}_\text{inter}\,\sigma\,\bigr]\;.
    \label{eq:falcon_interval_scaled}
\end{align}
\item Alternatively, we start with CQR, where the nonconformity measure of Eq.\eqref{eq:cqr_score} estimates the excess beyond the quantile edges via addition/subtraction. We adapt it via
\begin{align}
  R'_\text{CQR}(x) = R_\text{CQR}(x) - \tilde{q}_\text{inter}(x)\;.
  \label{eq:falcon_renorm_cqr}
\end{align}
Using $R'_\text{CQR}$ to construct $q_\alpha$ restores the marginal coverage guarantee of Eq.\eqref{eq:cp_sandwich} and gives
\begin{align}
    C_\alpha^{\text{FALCON-CQR}}(x)
    = \bigl[\,q_{\tau_{\text{lo}}} - \tilde{q}_\text{inter} - q_\alpha,\;
          q_{\tau_{\text{hi}}} + \tilde{q}_\text{inter} + q_\alpha\,\bigr]\;.
    \label{eq:falcon_interval_cqr}
\end{align}
\end{enumerate}
For constant $\tilde{q}_\text{inter}(x) = C$, the corrections can be absorbed into $q_\alpha$, restoring the confidence intervals of Eqs.\eqref{eq:cqr_interval} and \eqref{eq:prediction_set_scaled}.

\section{\texorpdfstring{$Z$}{Z} plus jets production}
\label{sec:results_amp_regression}

We benchmark the marginal and conditional conformal predictions for the LHC processes 
\begin{align} 
 q\bar{q} \to Z g (g) \; ,
\end{align}
for which the leading-order amplitudes have been encoded in surrogates before~\cite{Janssen:2023ahv,Herrmann:2025nnz, Bahl:2026jvt}. In particular, it has been shown that for the low final state multiplicity, the likelihood or posterior does not benefit from the high-dimensional central limit theorem and is hence non-Gaussian. This is why our standard methods do not provide calibrated uncertainties~\cite{Bahl:2026qaf}, and the amplitudes provide a perfect challenge for conformal predictions and their locally adaptive variants. Note that our numerical results differ slightly from~\cite{Bahl:2026qaf}, due to different training dataset sizes and the removed cut around small $|\cos\theta|$ for our analysis.

\begin{figure}[b!]
    \includegraphics[width=.45\linewidth]{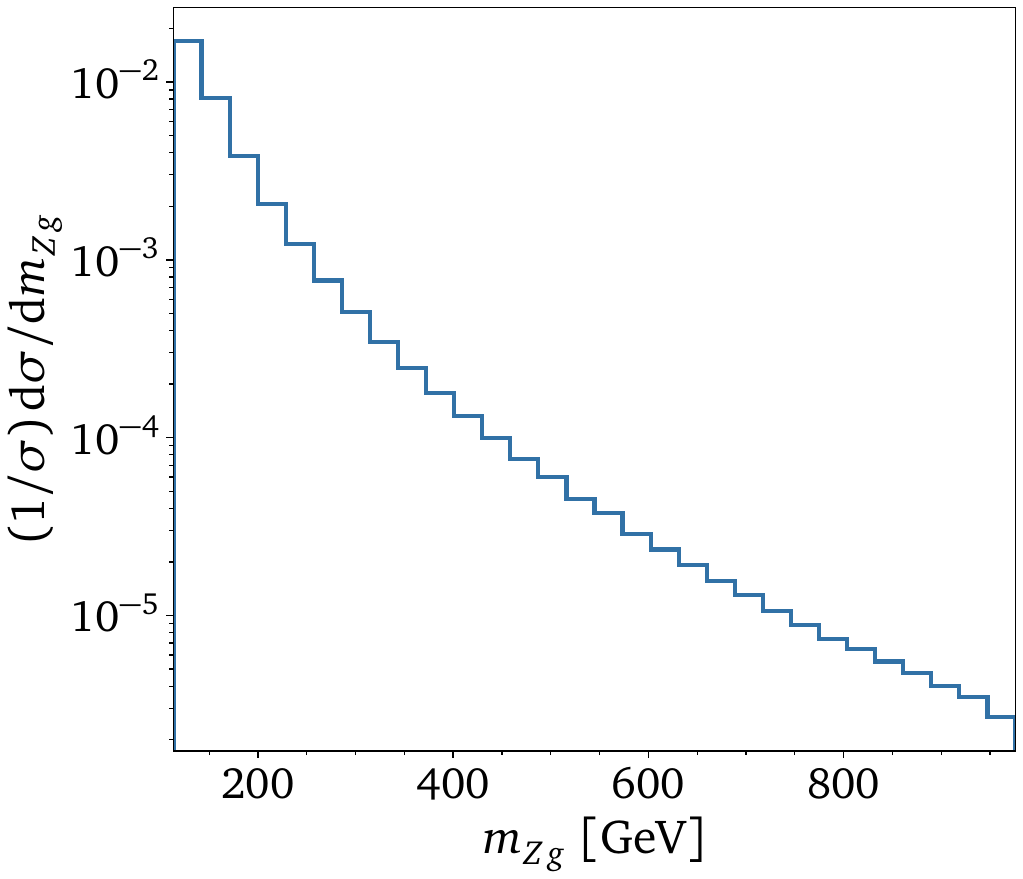}%
    \hfill
    \includegraphics[width=.45\linewidth]{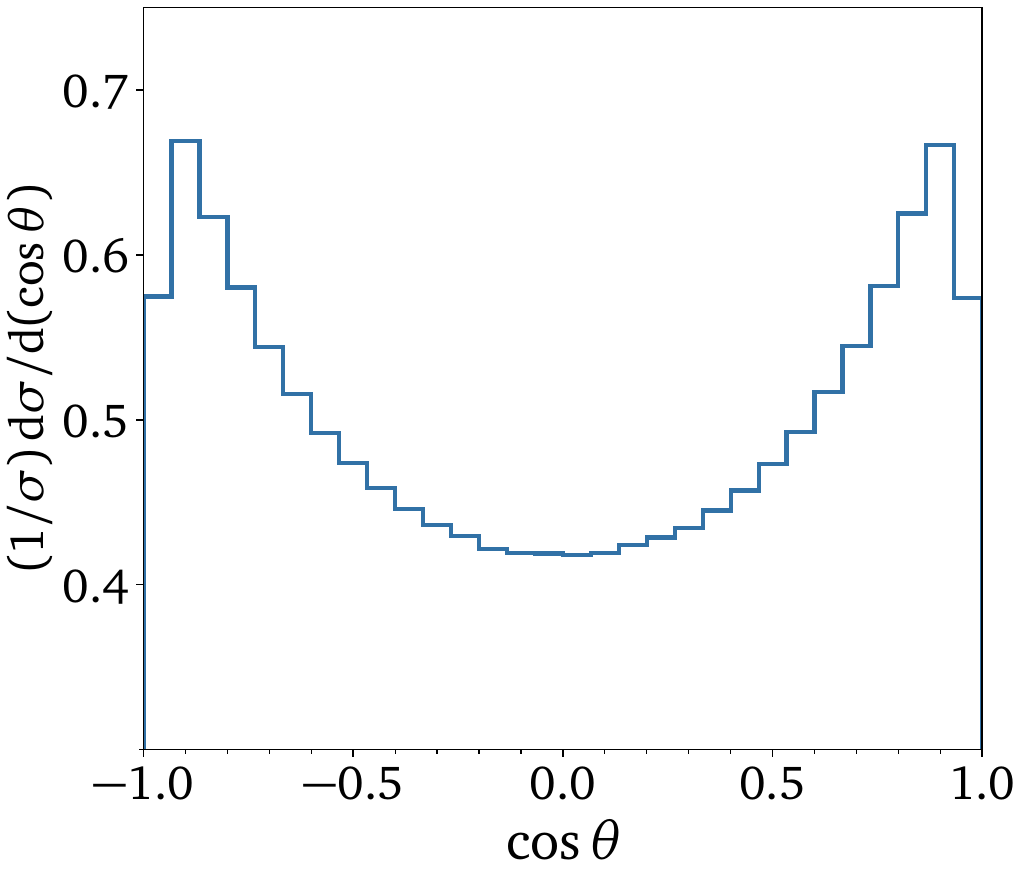}
    \caption{Invariant mass and scattering angle distribution for the $q\bar q\to Zg$ process.}
    \label{fig:Zg_distributions}
\end{figure}

The $Z g$ phase space is spanned by the invariant mass $m_{Zg}$ and the scattering angle $\cos\theta^*$. Including an additional gluon, we look at the invariant mass $m_{Zgg}$. We generate amplitudes with \madgraph using unweighted events, so the phase-space density follows the differential cross section. The only cuts we apply are $p_{T,g} >20~\gev$, and for the $2\to3$ process $\Delta R_{gg}>0.3$. The kinematic distributions for the $2 \to 2$ process are shown in Fig.~\ref{fig:Zg_distributions}.

\subsubsection*{Network and training}
\label{sec:implementation}

We first train surrogate networks using a heteroscedastic loss in Eq.\eqref{eq:het_loss} for an initial estimate of the amplitude $A_\text{NN}(x)$ and the uncertainty $\sigma(x)$. Then we use a conformal method to produce a calibrated prediction for the uncertainties. It predicts a set of 49 quantile predictions $q_\tau(x)$, the median with $\tau=0.5$ and 24 symmetric quantile pairs defined in Eq.\eqref{eq:qr_interval} and trained simultaneously via Eq.\eqref{eq:pinball_total}. Each surrogate is implemented as a five-layer MLP with 128 units per layer and GELU activations. They are trained with the Adam optimizer, using cosine-annealing. The detailed hyperparameters of the network are given in Tab.~\ref{tab:hyper}.

The training data for the surrogate network includes $3 \cdot 10^5$ points, split into training and validation datasets in a $2:1$ ratio. The validation dataset is used to avoid overfitting based on early stopping. In the training data, we augment the 4-vectors by Mandelstam log-invariants 
\begin{align}
  \log(p_i p_j) 
  \qquad \text{with on-shell 4-vectors} \qquad 
  (p_i p_j) \geq0 
  \; .
  \label{eq:log_invariants}
\end{align}
All inputs are standardized with training-set statistics. 

We evaluate the conformal methods on a separate dataset, which we generate like the training data but with a different random seed. We split it into a test set of 600k phase space points and a calibration pool of $n_\text{calib}=5\text{k}$ phase space points. The sensitivity to $n_\text{calib}$ is studied in Fig.~\ref{fig:n_calib_marginal_coverage}. We measure coverage on the test dataset, using Eq.\eqref{eq:marginal_coverage} for marginal coverage and Eq.\eqref{eq:bin_local_coverage} for conditional coverage. We summarize our five methods in Tab.~\ref{tab:methods}.

\begin{table}[t]
  \centering
  \renewcommand{\arraystretch}{1.2}
  \begin{small}
  \begin{tabular}{lllll}
  \toprule
Uncertainty & Method & Surrogate & Nonconformity Score & Interval \\
  \midrule
\multirow{1}{*}{Naive}
    & \textbf{CP}
    & Heteroscedastic
    & $R(x)$, Eq.\eqref{eq:test_residual}
    & Eq.\eqref{eq:cp_interval} \\[2mm]
\multirow{2}{*}{Adaptive}
&  \textbf{Het}
    & Heteroscedastic
    & $R_\sigma(x)$, Eq.\eqref{eq:scaled_residual_score}
    & Eq.\eqref{eq:prediction_set_scaled} \\
&  \textbf{CQR}
    & Quantile Regression
    & $R_\text{CQR}(x)$, Eq.\eqref{eq:cqr_score}
    & Eq.\eqref{eq:cqr_interval} \\[2mm]
\multirow{2}{*}{Local}
&  \textbf{FALCON-Het}
    & Heteroscedastic
    & $R_\sigma(x)$ + probe field
    & Eq.\eqref{eq:falcon_interval_scaled} \\
&  \textbf{FALCON-CQR}
    & Quantile Regression
    & $R_\text{CQR}(x)$ + probe field
    & Eq.\eqref{eq:falcon_interval_cqr} \\
  \bottomrule
\end{tabular}
\end{small}
  \caption{Conformal methods evaluated in this work, with the surrogate, nonconformity scores, and prediction interval used by each.}
  \label{tab:methods}
  \end{table}

We implement FALCON with 10 hotspot regions evenly spaced along $m_{Zg}$ with a width of 69~GeV. For each region, we generate 100 probe phase space points and evaluate the pre-trained surrogate on them to obtain the local conformal threshold $q_\alpha^{(i)}$. The interpolated field $\tilde{q}_\text{inter}(x)$ is then built by linear interpolation between the ten hotspot centers and their corresponding $q_\alpha^{(i)}$ values, assigning a smoothly varying local correction to every phase-space point. Since $\tilde{q}_\text{inter}(x)$ is fully determined by the probe phase space points, the calibration phase space points enter only as evaluation points at which the interpolant is queried — they play no role in shaping it — so the renormalized scores of Eq.\eqref{eq:falcon_renorm_scaled} and Eq.\eqref{eq:falcon_renorm_cqr} remain exchangeable. The global renormalization scalar $q_\alpha$ in Eq.\eqref{eq:falcon_interval_scaled} and Eq.\eqref{eq:falcon_interval_cqr}, computed from the same 5k calibration points used by the other conformal methods, then restores the marginal coverage guarantee. The choice of $n_\text{fresh}=100$ probe events per hotspot is validated in Fig.~\ref{fig:n_fresh_coverage} in the Appendix.

\subsection{Non-Gaussian uncertainties: \texorpdfstring{$q\bar{q}\to Zg$}{qq -> Zg}}
\label{sec:zg}

\begin{figure}[t]
    \centering
    \includegraphics[width=0.45\linewidth]{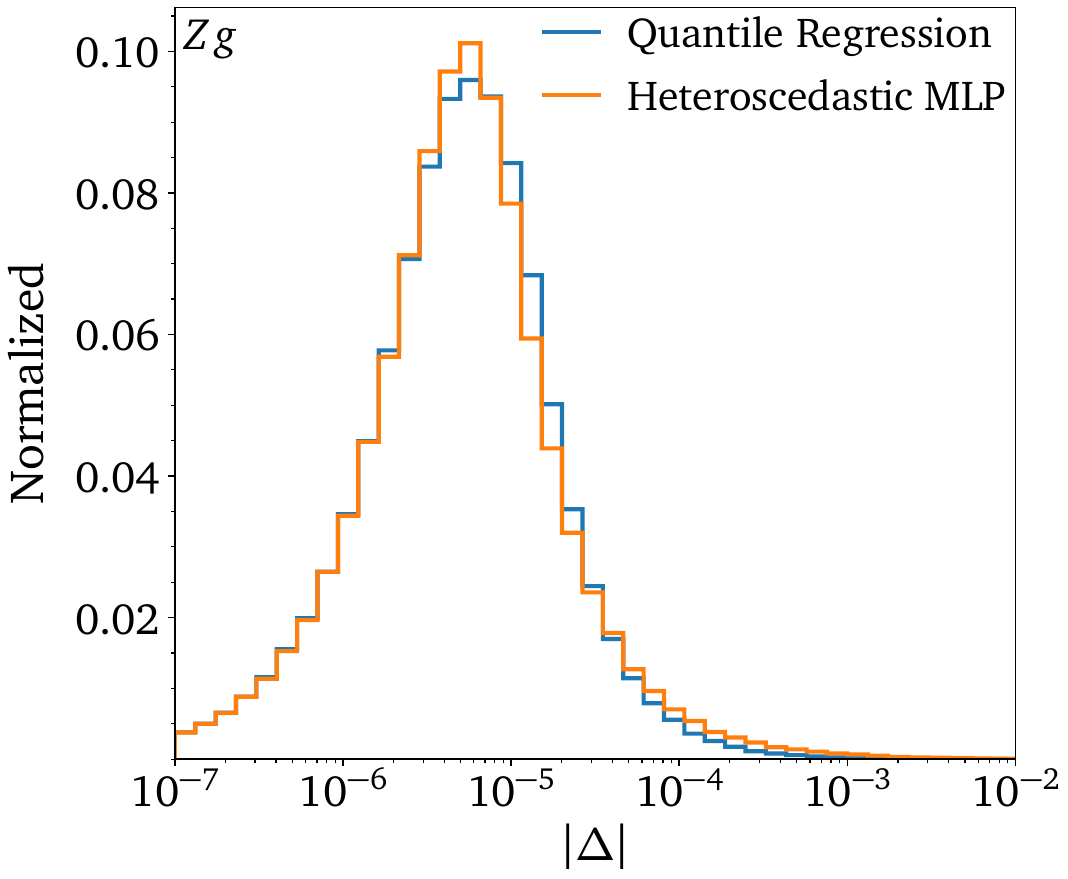}
    \caption{Accuracies of $\bar{q}q \to Zg$ amplitude surrogates trained with a heteroscedastic loss and with a pinball loss.}
    \label{fig: amplitude_accuracy_plot}
\end{figure}

We first quantify the performance of the surrogates in terms of the (relative) accuracy,
\begin{align}
    \Delta(x) = \frac{\text{A}_{\text{NN}}(x)-\text{A}_{\text{true}}(x)}{\text{A}_{\text{true}}(x)}\; .
    \label{relative accuracy}
\end{align}
For the Gaussian surrogate and its conformal variations, $A_{NN}$ corresponds to the predicted mean. For the quantile regression with Pinball Loss, $\text{A}_{\text{NN}}$ is given by the 0.5 quantile prediction corresponding to the median of the distribution. In Fig.~\ref{fig: amplitude_accuracy_plot} we show the accuracy of both approaches, both roughly following a Gaussian centered around $10^{-5}$. Towards smaller deviations, the tails look identical for both approaches, whereas towards larger deviations the quantile regression processing leads to a slightly more suppressed tail.

\subsubsection*{Marginal coverage}
\label{sec:zg_marginal}

\begin{figure}[t]
    \includegraphics[width=.44\linewidth]{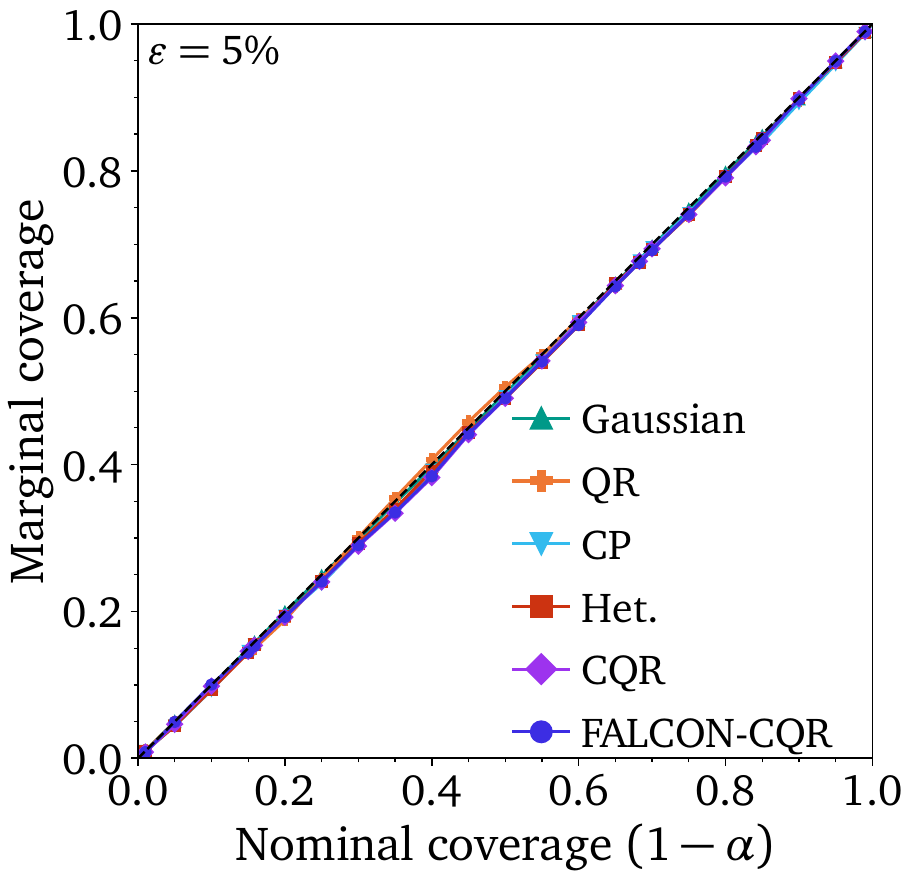}\hfill
    \includegraphics[width=.44\linewidth]{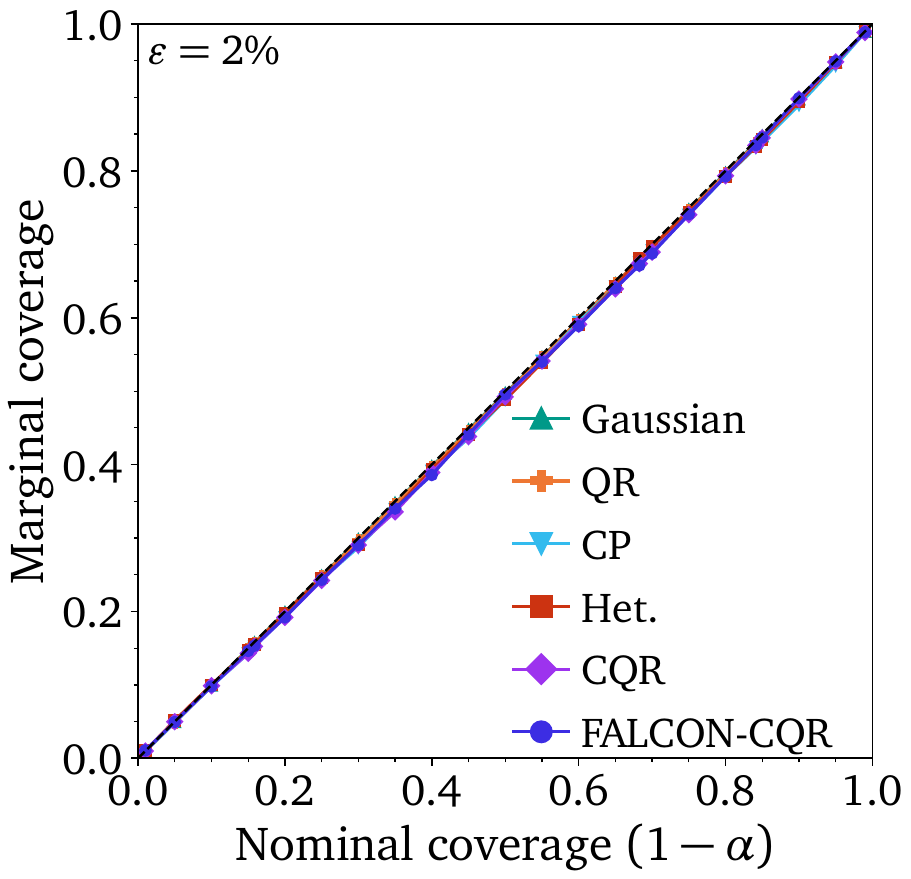} \\
    \includegraphics[width=.44\linewidth]{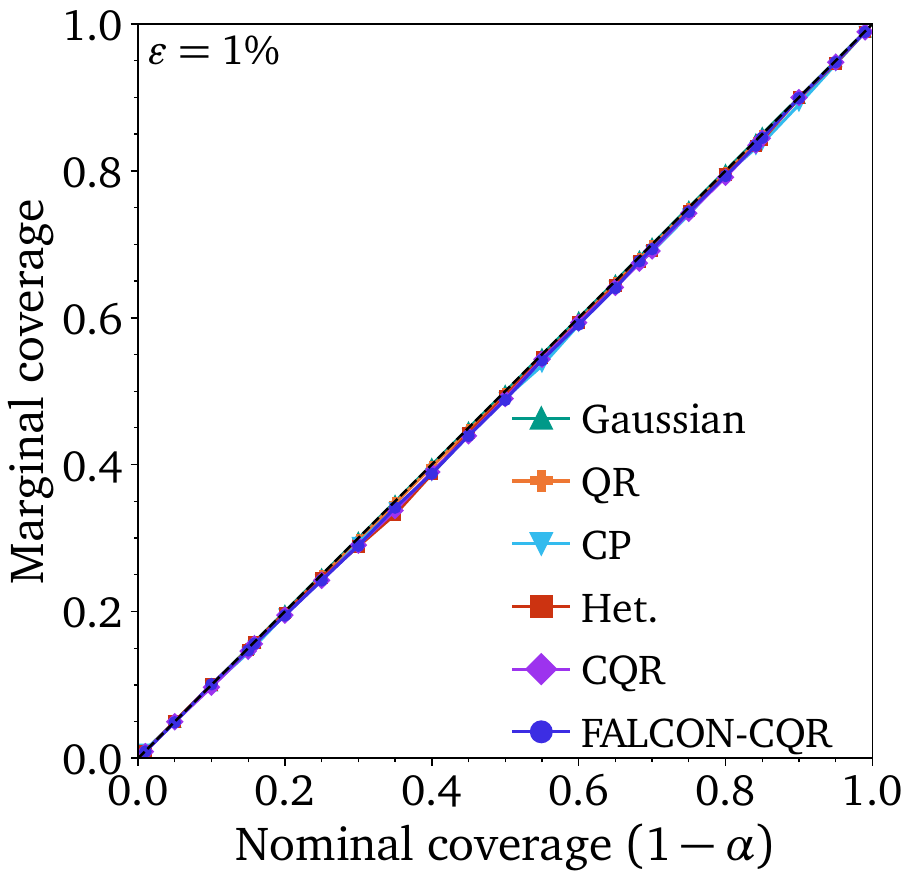}\hfill
    \includegraphics[width=.44\linewidth]{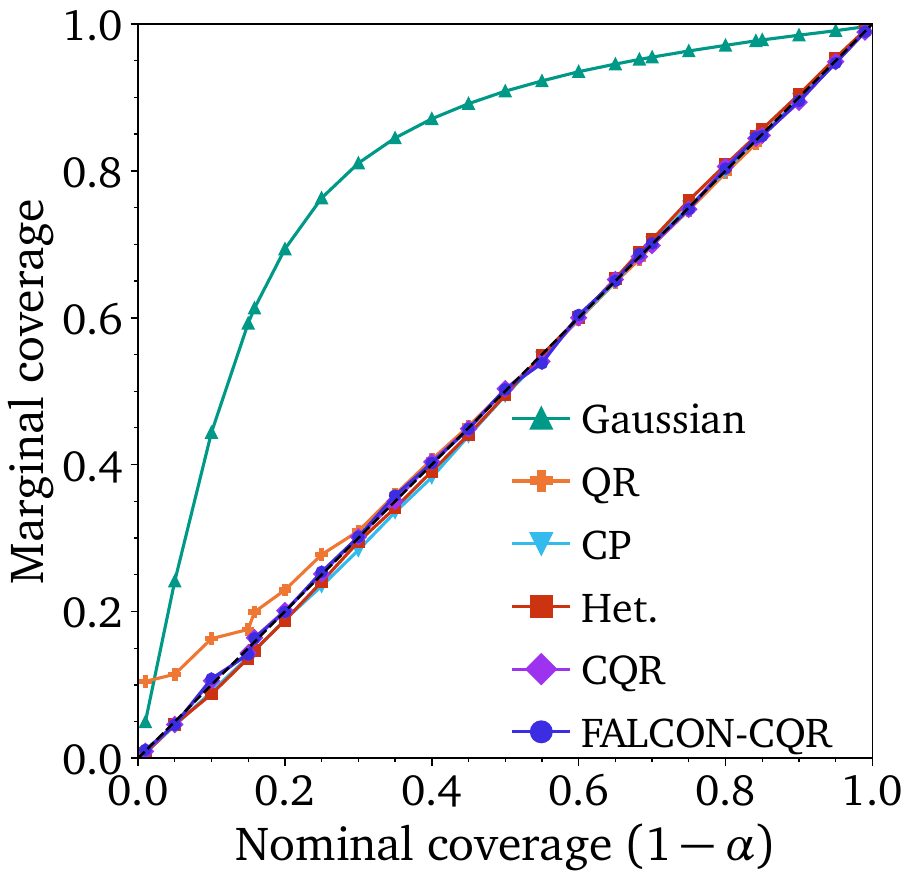}
    \caption{Marginal versus nominal coverage for decreasing levels of artificial Gaussian noise on the $q\bar{q}\to Zg$ training data.}
    \label{fig:marginal coverage}
\end{figure}

To test the coverage, we add Gaussian noise of 
\begin{align}
 \varepsilon = \left\{ 5\%, 2\%, 1\% \right\} 
\end{align}
to the training amplitudes. In Fig.~\ref{fig:marginal coverage} we compare the learned marginal coverage given by Eq.\eqref{eq:marginal_coverage} to the nominal coverage $1- \alpha$. A perfect global calibration would give a diagonal line. From 5\% down to 1\% noise, the marginal coverage is indeed perfectly calibrated for all methods.

Without artificial noise, the systematic uncertainties arise from limited phase space data, imperfect network, poor data representation, as well as noisy or imperfect training~\cite{Bahl:2025xvx}. From Fig.~\ref{fig: amplitude_accuracy_plot} we know that these uncertainties range from $10^{-3}$ to $10^{-5}$. Here, the marginal coverage of the heteroscedastic surrogate leads to a significant underconfidence of the amplitude precision. A Gaussian likelihood is not a valid assumption~\cite{Bahl:2025xvx}, \ie the standard deviation no longer corresponds to a 68\% confidence interval. This underconfidence is driven by outliers and enhanced tails in the probability distribution. Further details on the underlying distribution are given in App.~\ref{app:Pull_distribution}.

Quantile regression deviates from perfect calibration only for nominal coverage below 30\%, where instead of approaching zero it plateaus around 10\%. This is induced by fluctuations in the convergence of the quantile predictions. If we train the median quantile twice, we obtain two slightly different predictions $q_{0.5}(x)$ due to imperfect training. Evaluating the marginal coverage from the interval spanned by these two values explains the plateau because of the independent training of the quantile predictors. Going beyond the two surrogate networks, all conformal corrections lead to a perfect global calibration of the marginal coverage.

\subsubsection*{Conditional coverage}
\label{sec:zg_conditional}

\begin{figure}[t]
    \includegraphics[width=.45\textwidth]{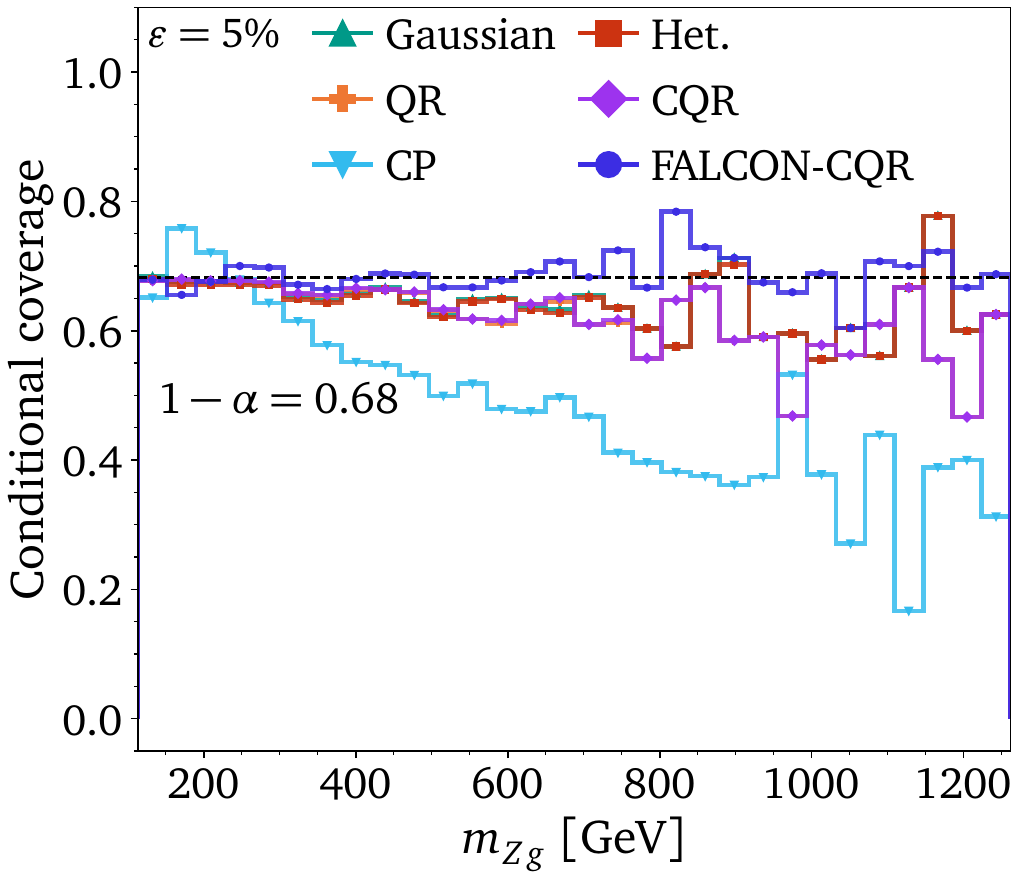}\hfill
    \includegraphics[width=.45\textwidth]{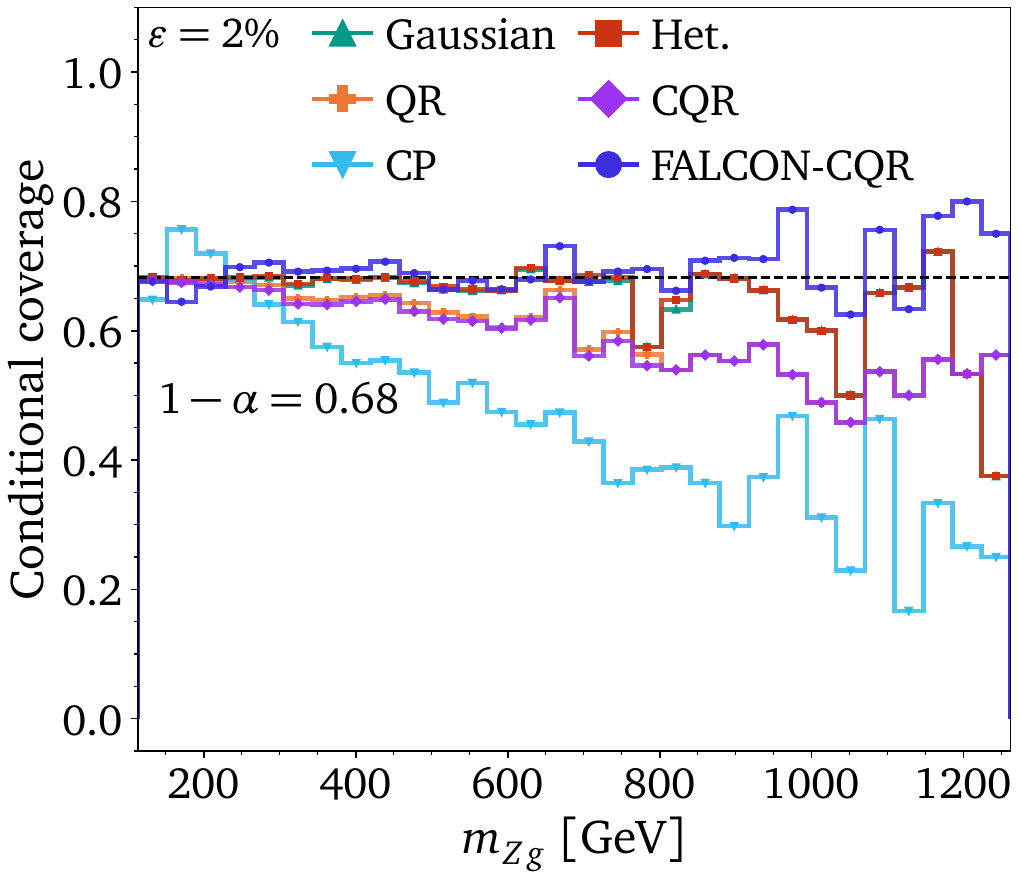} \\
    \includegraphics[width=.45\textwidth]{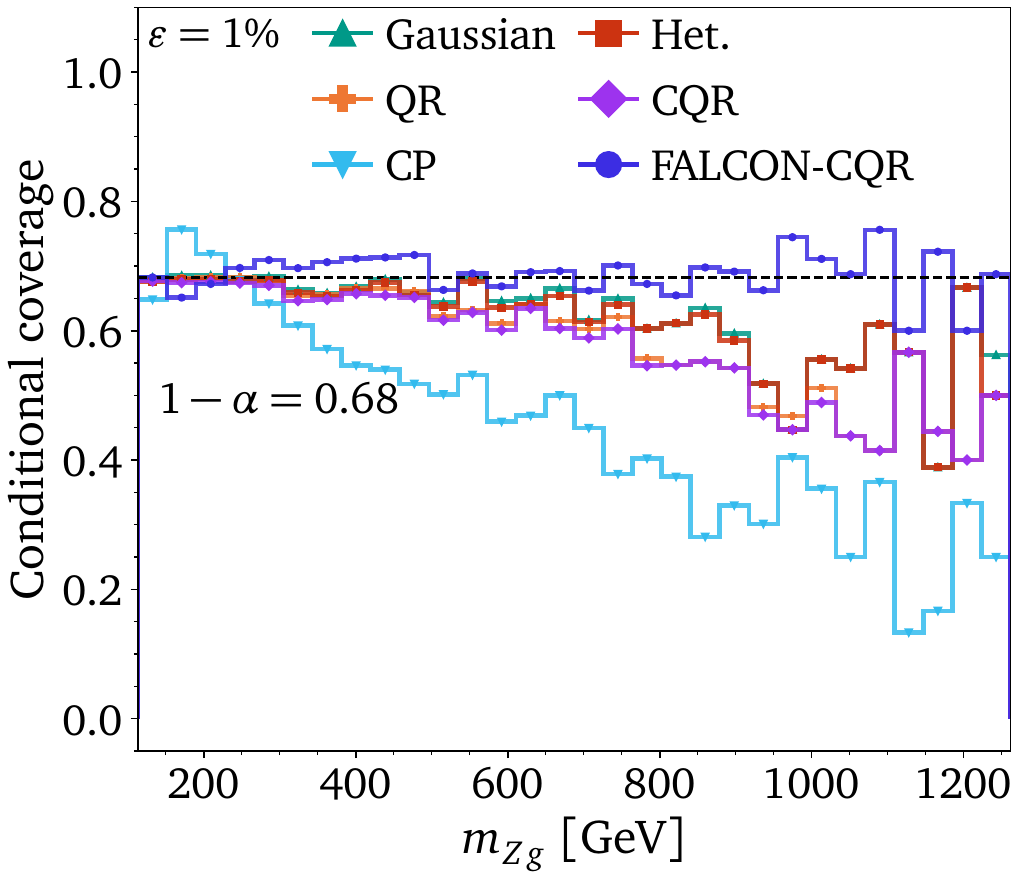}\hfill
    \includegraphics[width=.45\textwidth]{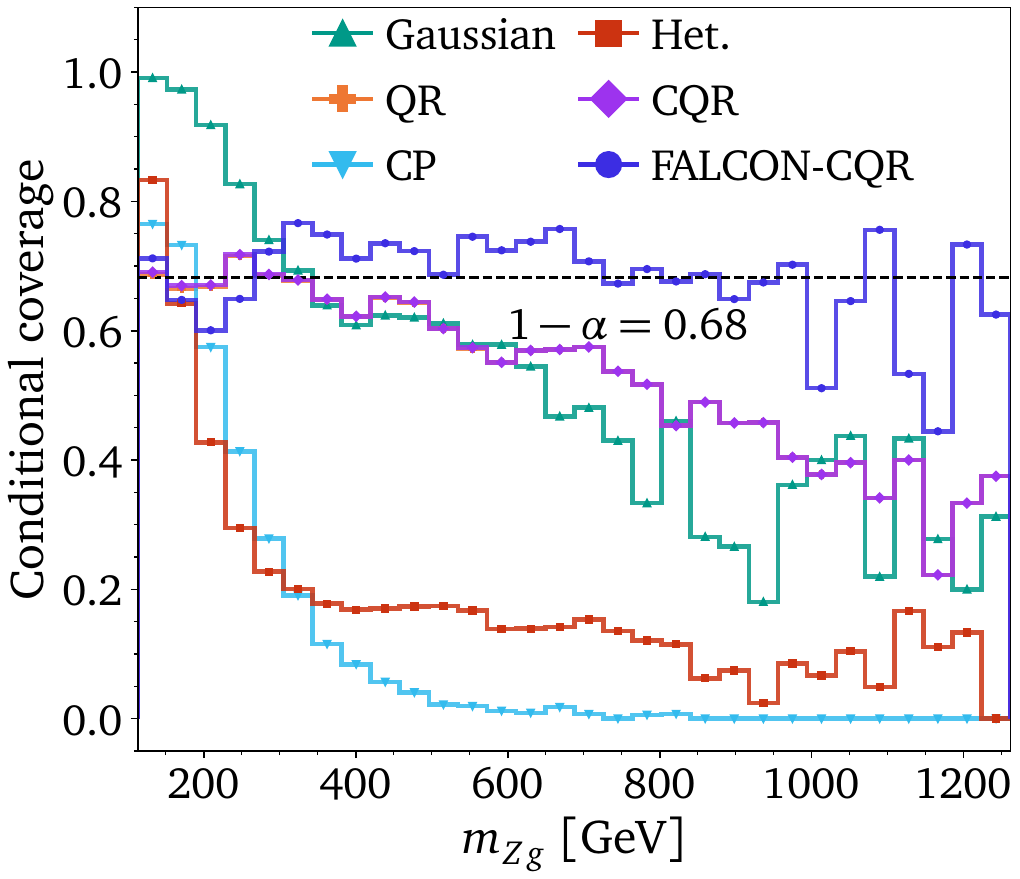}%
      \caption{Conditional 68\% CL coverage as a function of $m_{Zg}$, for decreasing artificial Gaussian noise. Coverage is evaluated at 10 evenly spaced probe windows.}
    \label{fig:invm_cqr_conditional_coverage}
\end{figure}

To test the local uncertainty prediction, we study the $m_{Zg}$ distribution with a reference value of 68\% or $1\sigma$ for the nominal coverage. As before, we add a decreasing amount of Gaussian noise. For the Gaussian and quantile regression surrogates, we find close to perfect conditional coverage for $m_{zg} \lesssim 400$~GeV. For larger invariant masses, where the artificial noise scales with larger amplitudes and the amount of training data is limited, we observe a slight overconfidence of roughly 5\%, implying that the surrogates do not expand their uncertainties sufficiently. Above 800~GeV, the limited statistics induce large fluctuations. Over the full $m_{Zg}$ range, corrections from the two adaptive methods are barely visible. For less noise, the overconfidence of the two surrogate networks in the high-mass tail becomes more pronounced.

Whereas naive and adaptive conformal predictions do not improve the local uncertainty estimate, our newly proposed FALCON method mitigates the overconfidence of the surrogates. It achieves near-nominal conditional coverage of $1-\alpha \approx 0.68$ across the entire kinematic range, until statistical limitations become dominant.

Without artificial noise, the low-mass region suffers from non-Gaussian behavior. The strong underconfidence in this bulk region explains the global underconfidence observed for the heteroscedastic surrogate before. For $m_{Zg} \gtrsim 400$~GeV, we observe increasing overconfidence, similar to the noisy case. Quantile regression behaves similarly for large invariant mass, due to the universal lack of training data. For the small invariant mass region, it provides substantially better coverage, as its more flexible network can adapt to the non-Gaussian uncertainties. The conformal predictions can correct the underconfidence in the bulk by globally reducing the size of the calibration intervals. This leads to a significant improvement at small invariant masses, but worsens the already observed overconfidence in the high-mass tail. Finally, the FALCON method improves the CQR calibration in the tail and reproduces the nominal coverage line across the full range. We stress-test FALCON against a sharply localised noise at a fixed threshold in App.~\ref{app:gpeak}.

\begin{figure}[b!]
    \includegraphics[width=.45\textwidth]{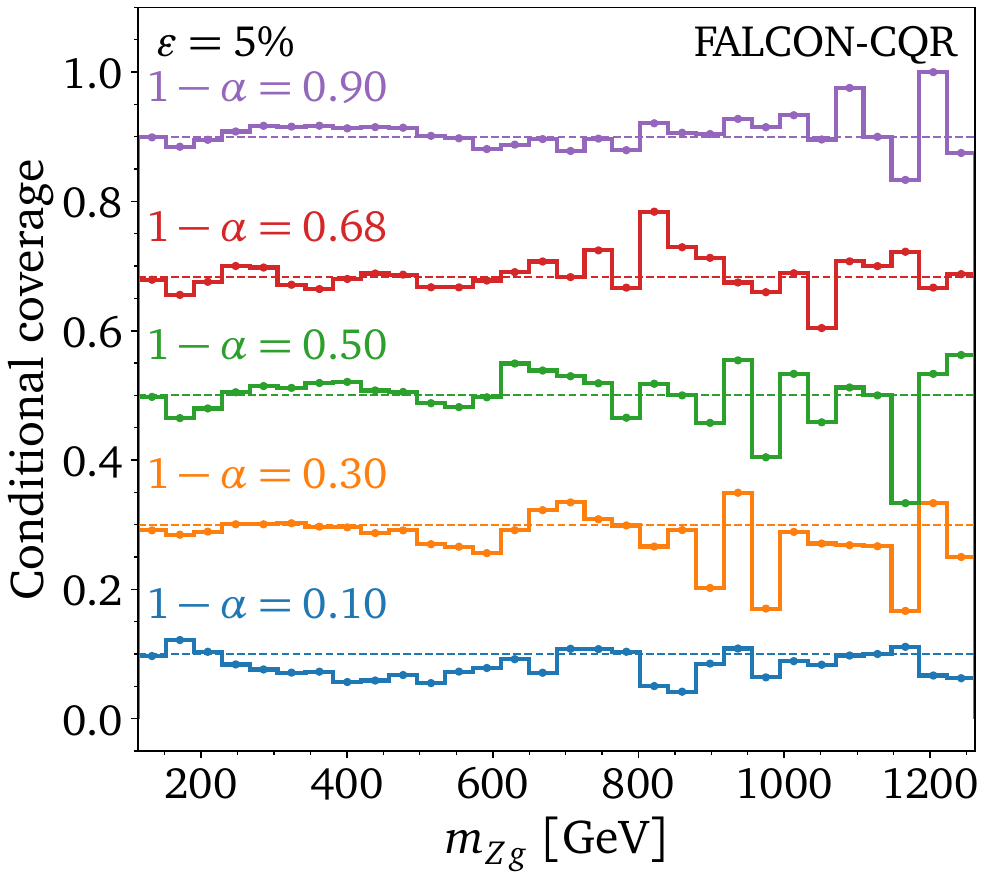}\hfill
    \includegraphics[width=.45\textwidth]{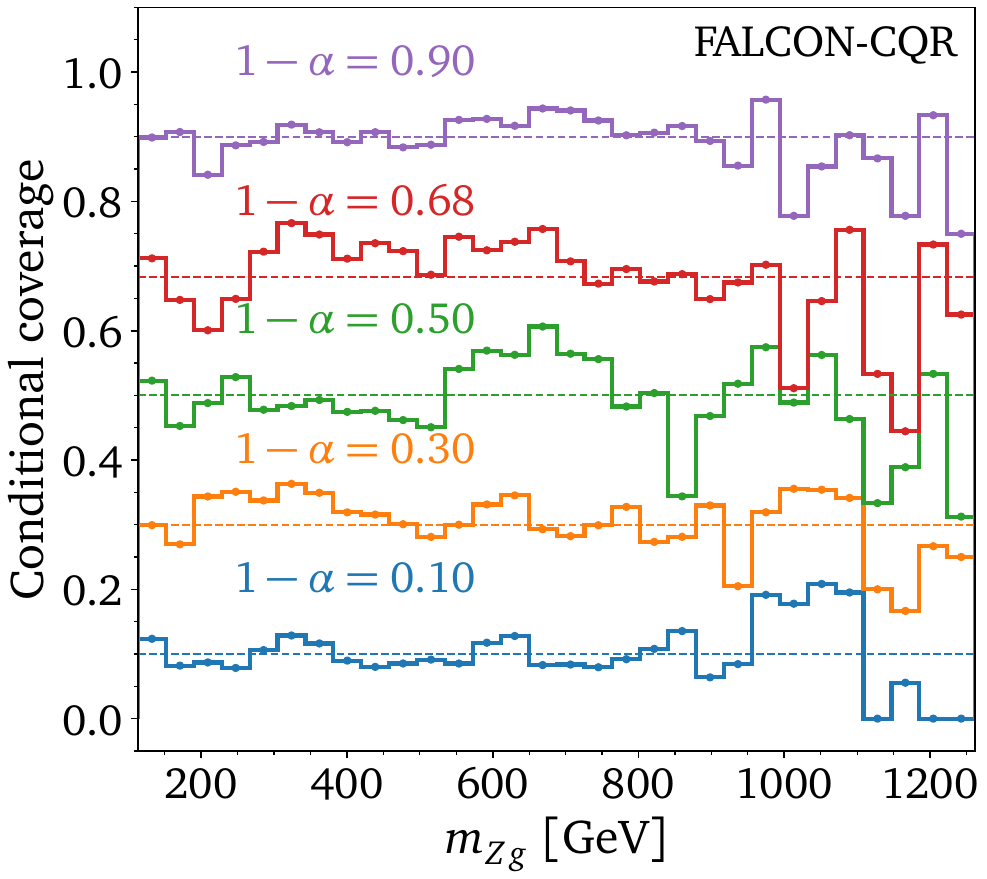}
    \caption{Conditional coverage from FALCON-CQR for different CLs as a function of $m_{Zg}$. Left: with artificial noise Right: without artificial noise.}
    \label{fig:conditional_coverage_vs_mzg}
\end{figure}

Given the success of FALCON for the 68\% CL, we confirm its robustness for confidence levels from 10\% to 90\% in Fig.~\ref{fig:conditional_coverage_vs_mzg}. The calibration curves show no systematic deviations,  neither in the bulk nor in the tail. Whereas the coverage is relatively stable below 500~GeV, fluctuations of the order of 10\% become dominant around 800~GeV. We further observe slightly increased fluctuations in the no-noise case for invariant masses below 400 GeV, related to the more complex structure of uncertainties in the absence of the Gaussian noise. 

\subsubsection*{Phase-space coverage}
\label{sec:zg_phasespace}

\begin{figure}[b!]
      \includegraphics[width=0.45\textwidth]{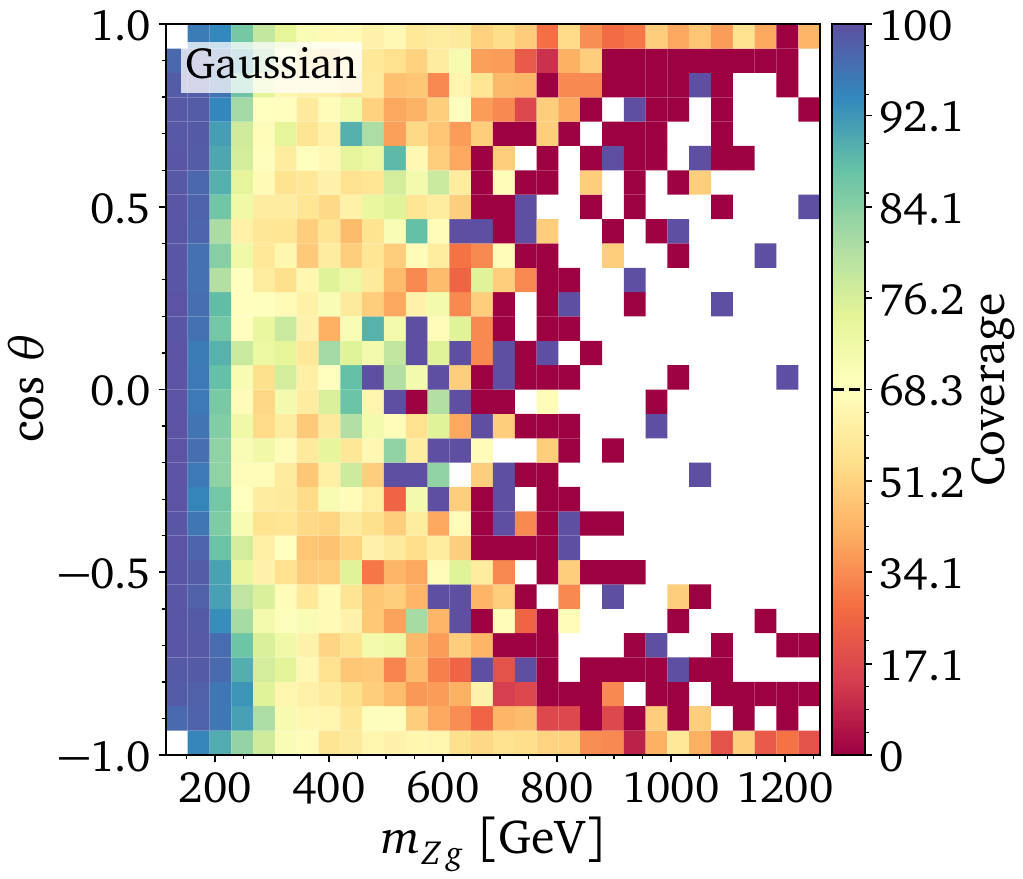} \hfill
      \includegraphics[width=0.45\textwidth]{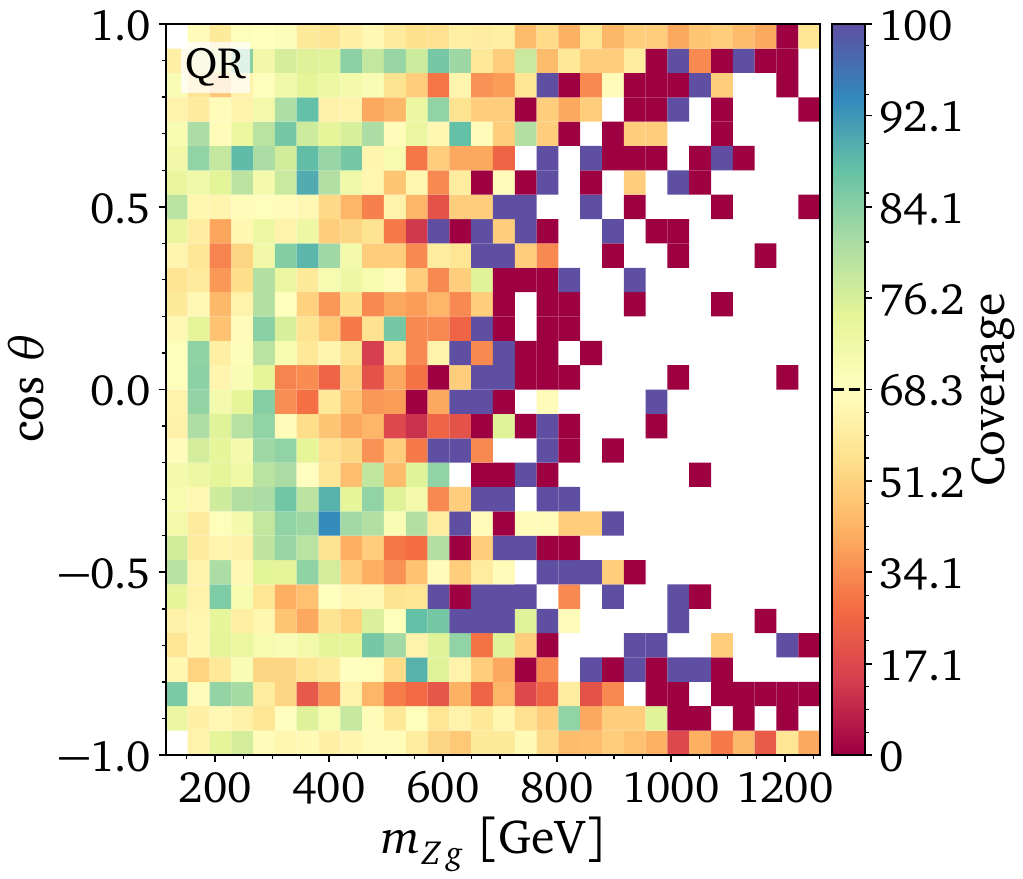} \\
      \includegraphics[width=0.45\textwidth]{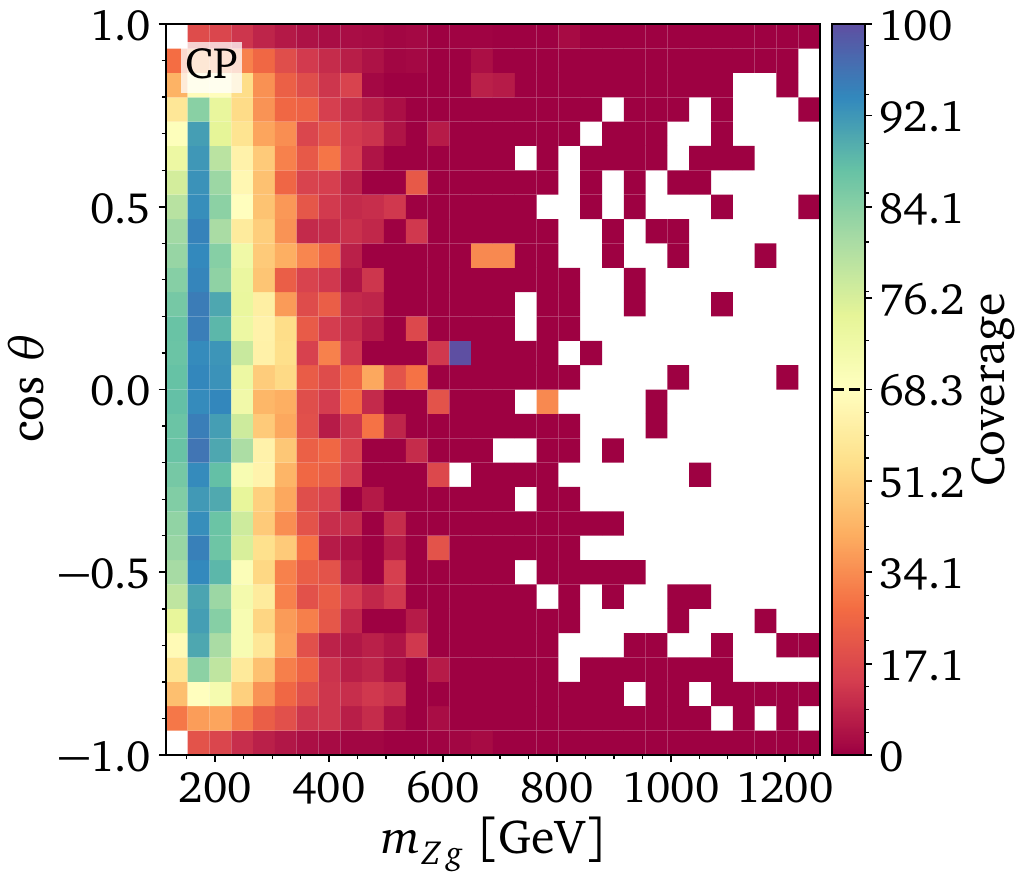} \hfill
      \includegraphics[width=0.45\textwidth]{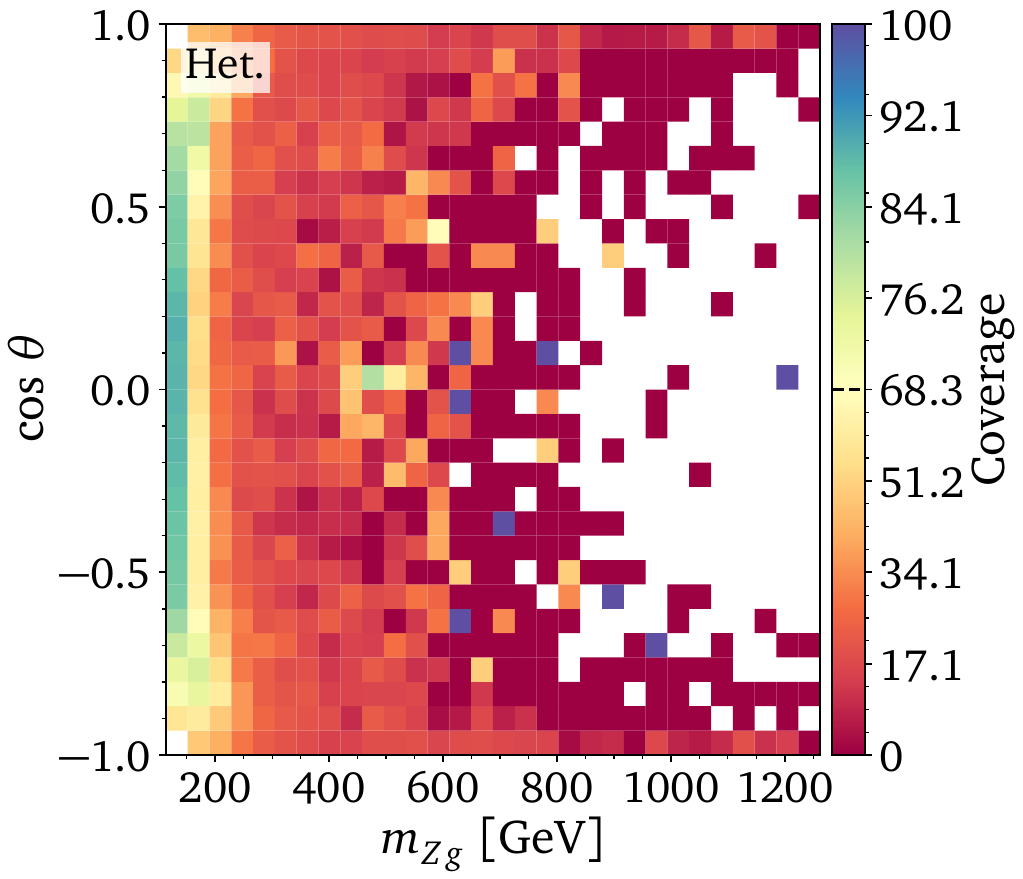} \\
      \includegraphics[width=0.45\textwidth]{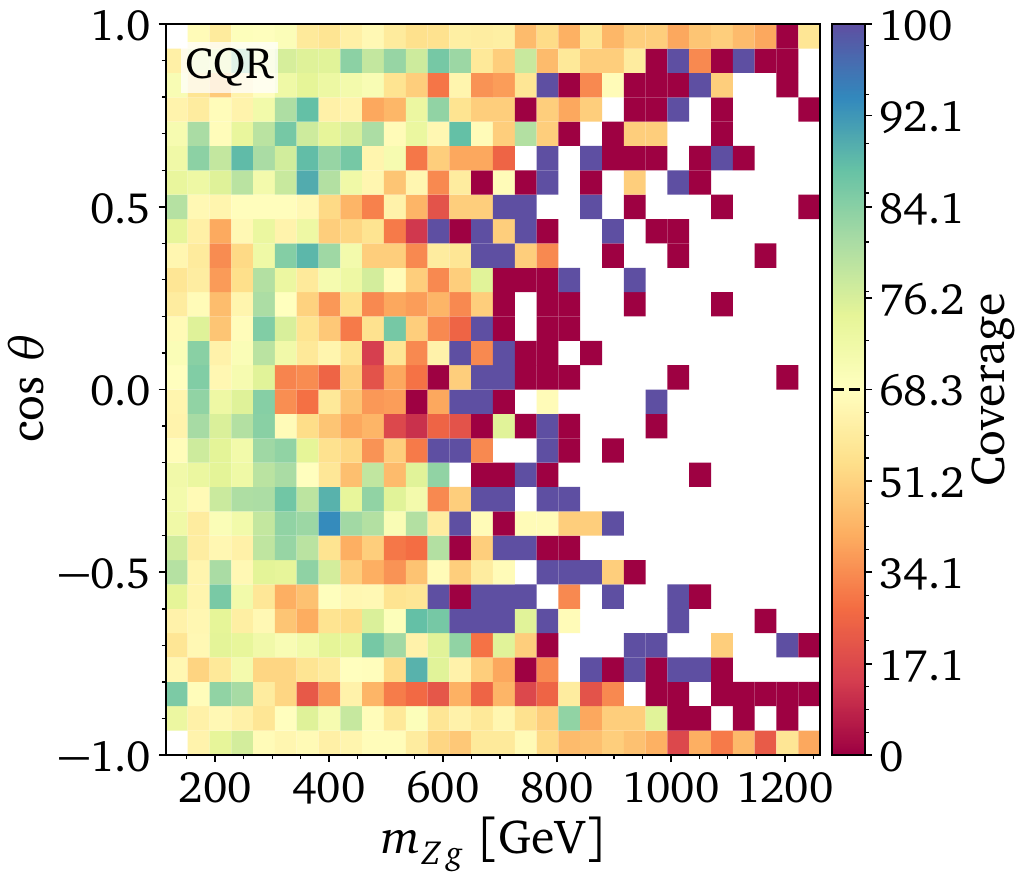} \hfill
      \includegraphics[width=0.45\textwidth]{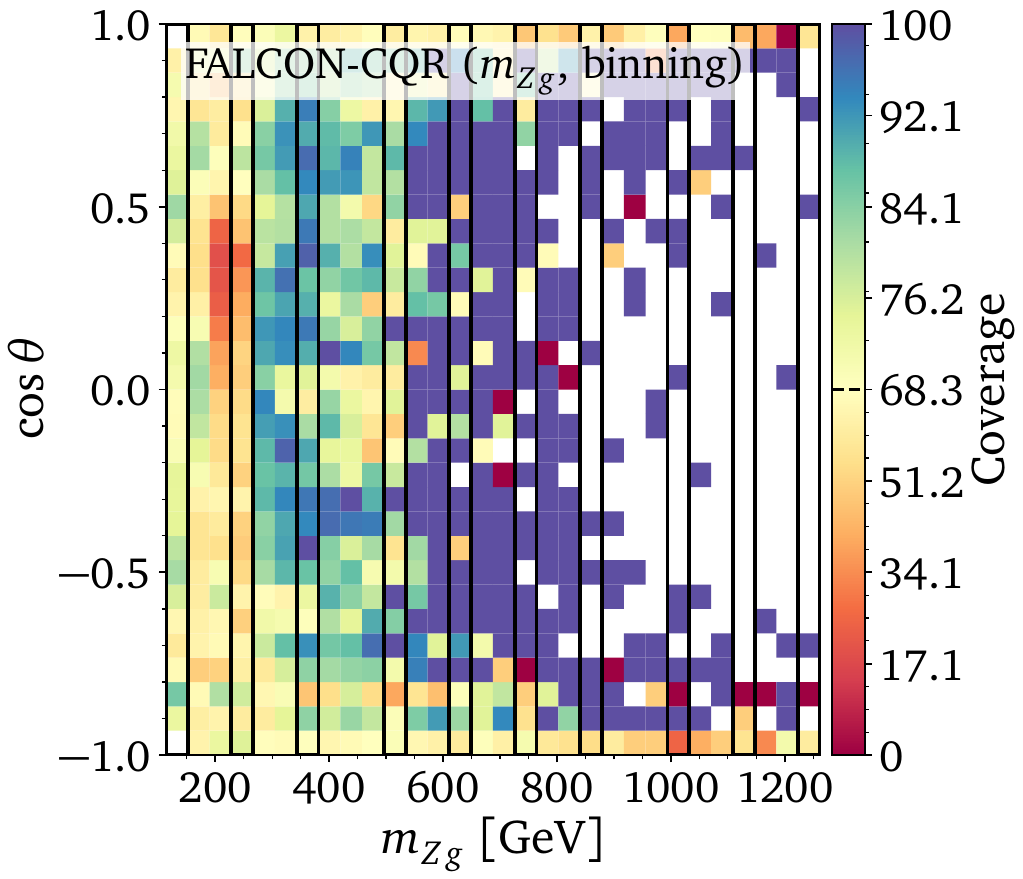}
\caption{Phase-space 68\% CL coverage in the $(m_{Zg},\cos\theta)$ plane, without artificial noise. The black lines in the FALCON panel indicate the probe
  windows at fixed $m_{Zg}$.}
  \label{fig:2d_local_coverage}
\end{figure}

Finally, we turn to the full physical $Zg$ phase space in Fig.~\ref{fig:2d_local_coverage}. The heteroscedastic Gaussian panel shows a three-zone structure driven primarily by $m_{Zg}$. For $m_{Zg} \lesssim 200$~GeV the Gaussian assumption breaks down and the coverage reaches 100\% uniformly across the full $\cos\theta$ range. At moderate invariant masses the coverage recovers near-nominal values, with little impact along $\cos\theta$. A dependence on $\cos \theta$ appears for $m_{Zg} > 600$~GeV, where the coverage drops below 50\% as a function of $m_{Zg}$, but in the collinear regime with its larger statistics it remains more stable.

The naive and adaptive conformal predictions behave similarly, the only difference originating from the global reduction of the coverage to achieve marginal coverage across $m_{Zg}$. For small $m_{Zg}$, the collinear regions are actually undercovering the confidence intervals. Quantile regression achieves near-nominal coverage across the bulk of the low-to-moderate mass range, as expected from the mass-dependent conditional coverage. In the high-mass tail, the coverage becomes noisy and locally mis-calibrated, eventually dropping to almost zero with large bin-to-bin fluctuations. The local coverage structure of CQR is inherited entirely from the underlying quantile regression.

The last panel of Fig.~\ref{fig:2d_local_coverage} shows the FALCON coverage. The black lines indicate the probe windows that span the full $\cos\theta$ range for fixed $m_{Zg}$. The correction $\tilde{q}_\text{inter}$ captures the $m_{Zg}$ dependence of the nonconformity measure but is blind to the residual $\cos\theta$ dependence. The 1D conditional coverage of Fig.~\ref{fig:invm_cqr_conditional_coverage} shows FALCON to be close to the nominal 68\% across the full invariant mass range. At low mass, coverage is precisely measured and compatible with the nominal value. The overconfidence region around $m_{Zg} \approx 150~...~300$~GeV and $\cos\theta \approx 0.3~...~0.6$ is inherited from the underlying quantile regression network and appears identically in the CQR and QR panels. Above $m_{Zg} \gtrsim 500$~GeV, the calibration is dominated by the collinear regime, where FALCON recovers near-nominal coverage. Outside the collinear region, sparse training data leads to a patchwork that is dominated by underconfidence, with occasional bins reflecting the complementary outcome. Compared to all other methods, FALCON uniquely maintains near-nominal coverage across phase space. The coverage along $\cos\theta$ is shown in Fig.~\ref{fig:zg_invm_interpolation_costheta}; the 1D probe binning along $m_{Zg}$ does not degrade the angular calibration. An alternative 2D probe scheme is explored in Fig.~\ref{fig:falcon_2d} and performs less well, since the underlying QR surrogate already captures the $\cos\theta$ dependence.

\subsection{Higher-dimensional phase space: \texorpdfstring{$q\bar{q}\to Zgg$}{qq -> Zgg}}
\label{sec:zgg}

\begin{figure}[b!]
    \includegraphics[width=0.45\linewidth]{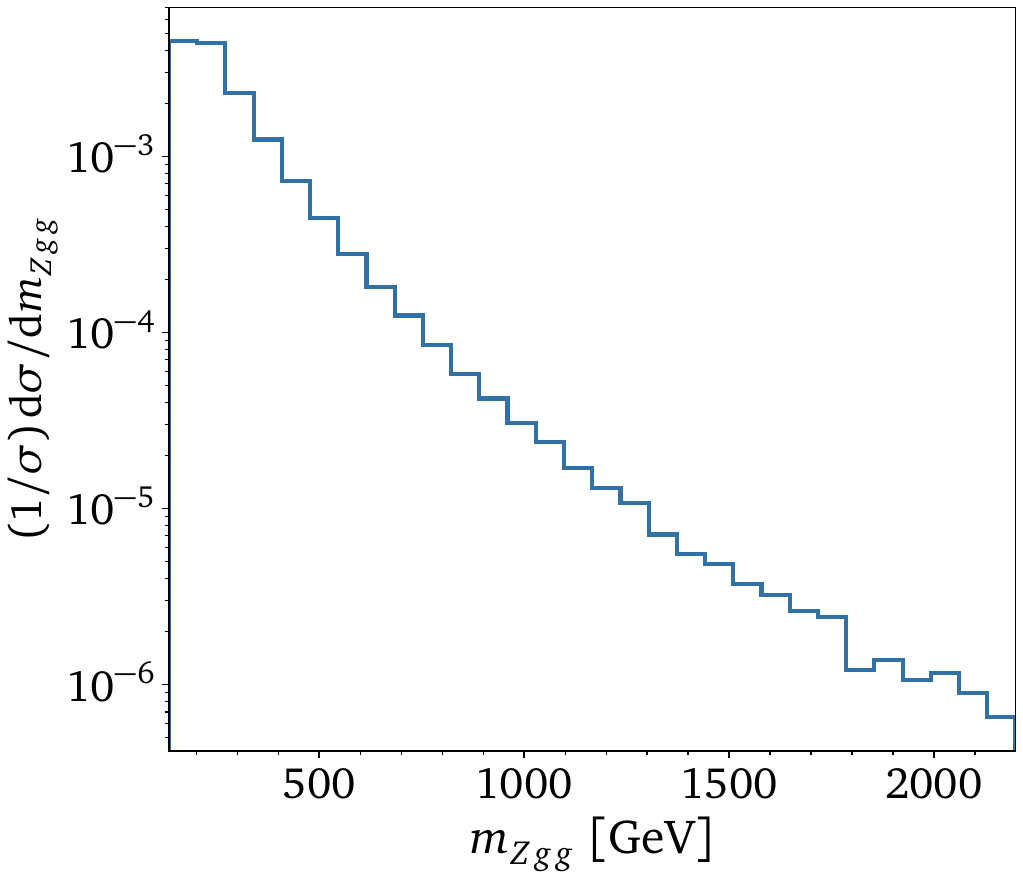}%
    \hfill
    \includegraphics[width=0.45\linewidth]{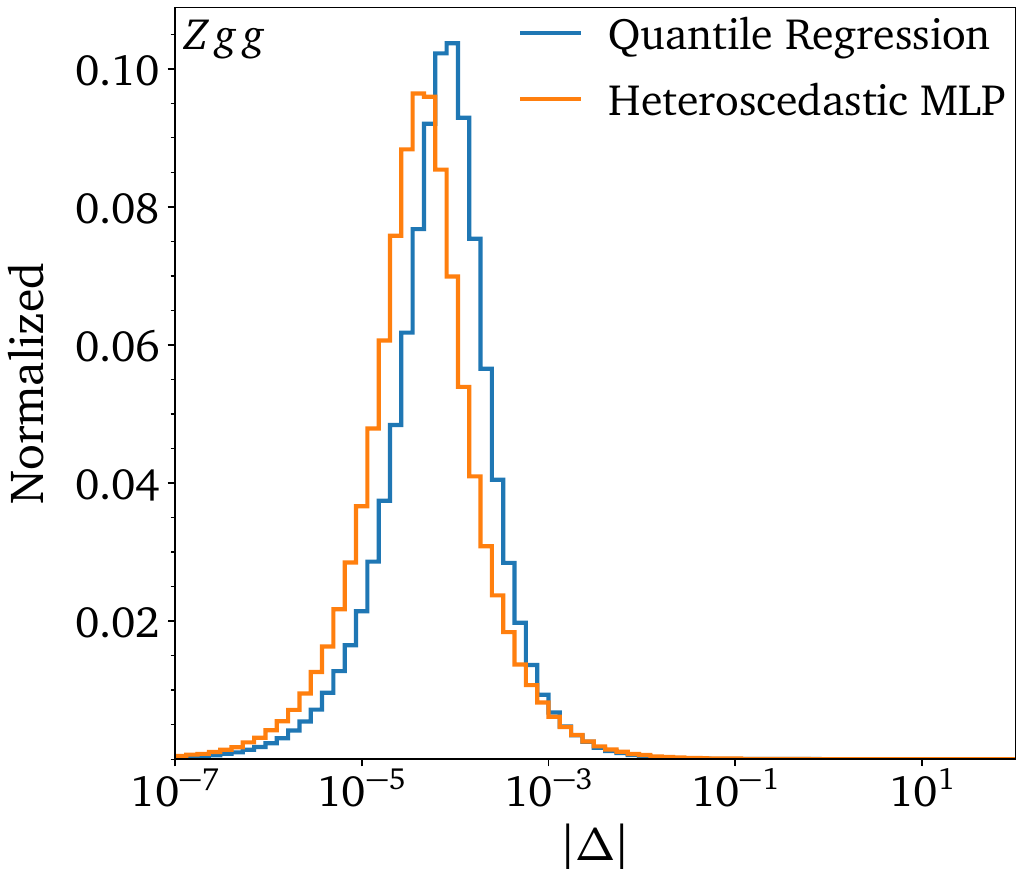}
    \caption{Left: invariant mass distribution $m_{Zgg}$ of $q\bar{q}\to Zgg$. Right: relative accuracy of the $q\bar{q}\to Zgg$ amplitude surrogates trained with a heteroscedastic loss and with a pinball loss.}
    \label{fig:zgg_dist_accuracy}
\end{figure}

When residuals decompose as a sum of independent contributions, the central limit theorem predicts that their distribution becomes increasingly Gaussian as the number of phase-space dimensions grows~\cite{Bahl:2026qaf}. This raises the question if we can scale conformal predictions to the process $q \bar q \to Zgg$, before the systematics becomes Gaussian for higher-multiplicity final states. 

We start by illustrating the invariant mass distribution $m_{Zgg}$ in the left panel of Fig.~\ref{fig:zgg_dist_accuracy}. The surrogate accuracies for the heteroscedastic Gaussian approach and the quantile regression are shown in the right panel, with peak values closer to $\Delta \approx 10^{-4}$, almost one order of magnitude worse than the $Zg$ results. This is a result of the higher dimensionality and complexity of the phase space and could, if needed, be recovered with more training data~\cite{Bahl:2026jvt}. 

As for $q \bar q \to Zg$, both surrogates are accurate across the invariant mass range, following a Gaussian loosely. While for $Zg$ the two approaches perform almost identically, the peak of the QR surrogate is now shifted toward slightly larger errors relative to the heteroscedastic surrogate, indicating already that a Gaussian description might be more suitable. Moreover, the tail towards vanishing deviations shows a stronger suppression for the QR surrogate. 

\begin{figure}[t]
    \includegraphics[width=0.45\linewidth]{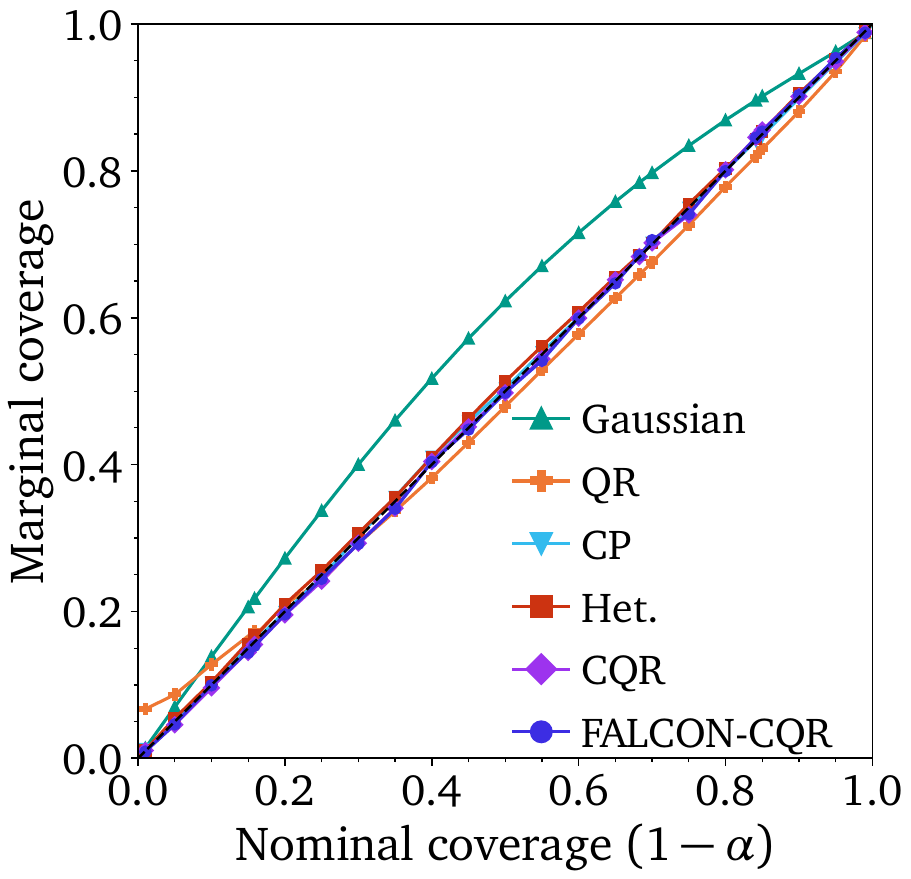}
    \hfill
    \includegraphics[width=0.485\linewidth]{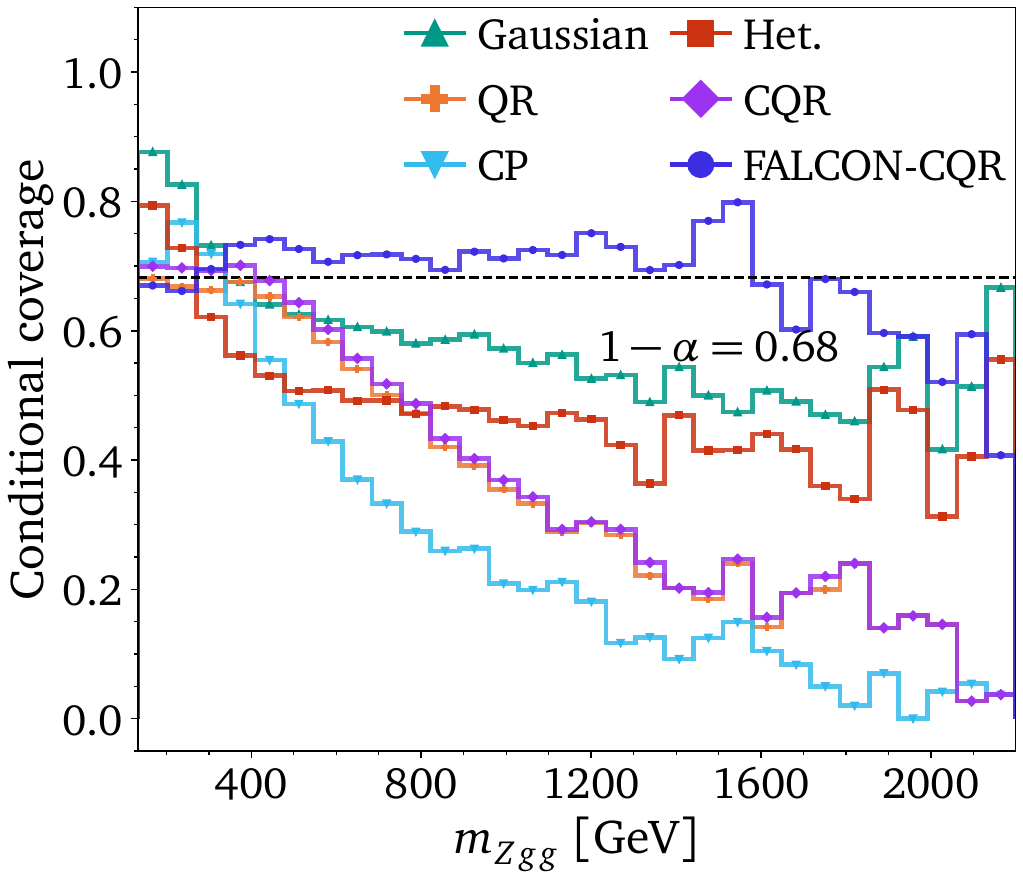}%
    \caption{Left: marginal versus nominal coverage for $q\bar{q}\to Zgg$. Right: conditional 68\% CL coverage as a function of $m_{Zgg}$, evaluated at 10 evenly spaced probe windows.}
    \label{fig:zgg_coverage_plot_cqr}
\end{figure}

The marginal coverage in the left panel of Fig.~\ref{fig:zgg_coverage_plot_cqr} reproduces a similar pattern as for $Zg$. The two surrogate networks display a distinct underconfidence, in particular the heteroscedastic Gaussian. However, compared to $Zg$ production, the degree of underconfidence is visibly reduced, consistent with more Gaussian residuals. Quantile regression slightly overcovers towards vanishing coverage, due to fluctuations between quantiles which prevent a convergence to zero. By construction, all conformal methods are well-calibrated.

The conditional coverage as a function of $m_{Zgg}$ is shown in the right panel of Fig.~\ref{fig:zgg_coverage_plot_cqr}. The most striking difference with respect to $Zg$ production is the Gaussian baseline. While the coverage remains below the target value of 68\% for large invariant masses, it stabilizes between 50\% and 60\%, starting to fluctuate strongly around 1.8~TeV. Again, this improvement is expected for the higher dimensionality because of the central limit theorem. In the low-mass bulk, the Gaussian approach is still uncerconfident, albeit less severely than for $Zg$. Quantile regression achieves nominal coverage in the low-mass bulk region, which plays to its flexibility. Towards higher $m_{Zgg}$ it drops below 20\% coverage. 

The corrections from naive and adaptive conformal predictions target the low-mass area and lead to a reduced size of the confidence intervals. Consequently, the coverage drops consistently over the full phase space. The large variations in the amplitude values make the scaled approach more suitable. Finally, FALCON follows the nominal coverage closely. Throughout most of the $m_{Zgg}$ range, FALCON exceeds the nominal coverage by less than 10\%, until the evaluation is dominated by statistical fluctuations.

\begin{figure}[t]
    \includegraphics[width=.48\textwidth]{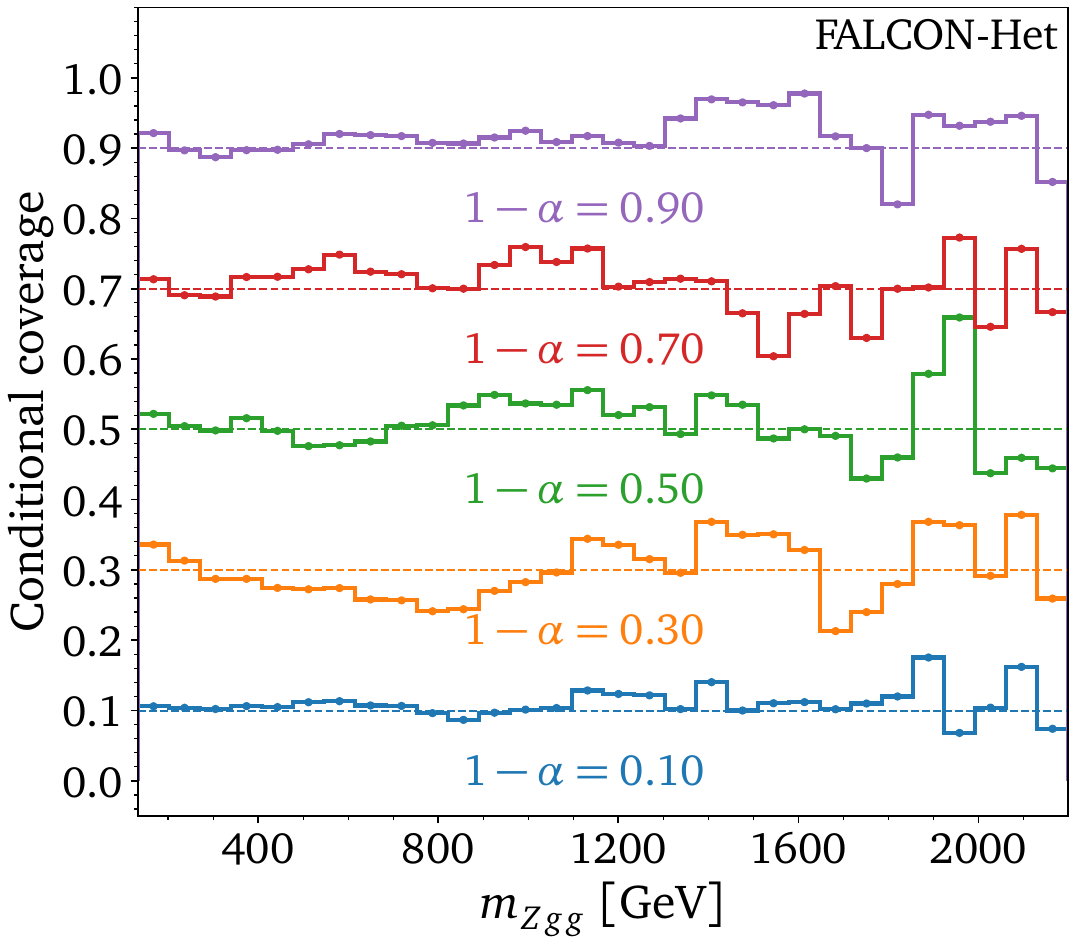}\hfill
    \includegraphics[width=.48\textwidth]{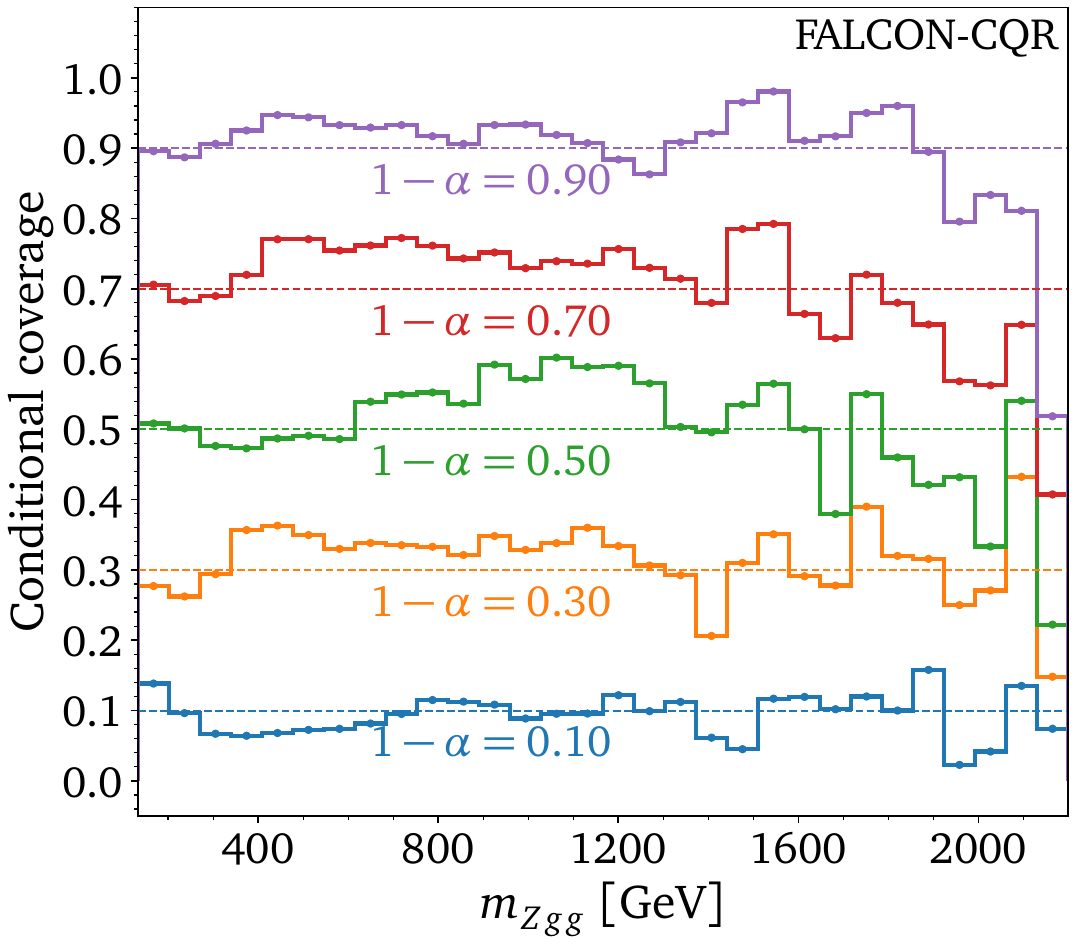}%
    \caption{Conditional coverage from FALCON for different CLs as a function of $m_{Zgg}$ without artificial noise. Left: FALCON-Het. Right: FALCON-CQR.}
    \label{fig:conditional_cov_vs_mzgg}
\end{figure}

Since for higher dimensionality the Gaussian approach outperforms quantile regression, we compare the two FALCON versions for different nominal coverage levels in Fig.~\ref{fig:conditional_cov_vs_mzgg}. FALCON-Het tracks the nominal coverages more closely at all levels. This is consistent with the different accuracy peaks of the underlying surrogates. The heteroscedastic surrogate captures the $Zgg$ bulk more accurately than the QR surrogate, consistent with increasingly Gaussian uncertainties. In both cases, FALCON maintains near-perfect calibration across the full $m_{Zgg}$ range, confirming that its approach is effective also for higher-dimensional phase space. 

\section{Outlook}
\label{sec:conclusions}

Precise amplitude surrogates with calibrated uncertainties are an essential ingredient to ultra-fast event generation for the LHC. While a heteroscedastic loss, Bayesian networks, and evidential regression have shown excellent performance for moderate to large final state multiplicities, loop amplitudes for two or three final-state particles remain a challenge. The reason is that the established methods for learning calibrated uncertainties require us to model the underlying systematics. For low-dimensional final states, these systematics cannot be treated Gaussian-like, which means the learned uncertainties are poorly calibrated.  

In this paper, we have targeted this problem with quantile regression and conformal prediction. These methods are distribution-free and can be applied as a post-processing to the surrogate training. We have focused on two toy models, $Z$-production with one or two gluons. These simple amplitudes allow us to test our new method without the numerical cost of actual higher-loop amplitude computations~\cite{Breso-Pla:2024pda,Bahl:2026qaf}.

For the $Zg$ amplitude with a relative accuracy of $10^{-5}$, a heteroscedastic loss fails to provide calibrated uncertainties. Naive and adaptive conformal predictions provide the correct coverage, but neither of them encodes localized effects in phase space. Our new FALCON method can be applied to a heteroscedastic or quantile regression surrogate, using the heteroscedastic residual or CQR as the nonconformity measure. In both cases, the FALCON interpolation between hotspot regions provides us with locally calibrated uncertainties across phase space, even for highly non-Gaussian systematics.

For the $Zgg$ amplitude, the performance of the heteroscedastic loss alone improves, but the learned uncertainties remain poorly calibrated. Again, FALCON provides us with correctly calibrated uncertainties over the entire phase space. Going to higher-dimensional final states, it will be limited by the number of necessary hotspot regions, but given that the systematics will become more Gaussian, our new FALCON methods perfectly complement existing methods and allow us to learn calibrated uncertainties for LHC processes for any number of particles in the final state.

\subsection*{Acknowledgements}

We are very grateful to Pascal Memmesheimer for fruitful discussions. This work is supported by the Deutsche Forschungsgemeinschaft (DFG, German Research Foundation) under grant 396021762 -- TRR~257 \textsl{Particle Physics Phenomenology after the Higgs Discovery} and funded by the Carl-Zeiss-Stiftung through the project \textsl{Model-Based AI: Physical Models and Deep Learning for Imaging and Cancer Treatment}. We have received funding from the European Union’s Horizon Europe research and innovation programme under the Marie Sklodowska-Curie grant agreement No 101168829, \textsl{Challenging AI with Challenges from Physics: How to solve fundamental problems in Physics by AI and vice versa} (AIPHY). The authors acknowledge support by the state of Baden-Württemberg through bwHPC and the German Research Foundation (DFG) through grant no INST 39/963-1 FUGG (bwForCluster NEMO). AB gratefully acknowledges the continuous support from LPNHE, CNRS/IN2P3, Sorbonne Université and Université de Paris Cité.

\appendix
\clearpage
\section{Hyperparameters}
\label{app:hyper}

\begin{table}[h]
  \centering
  \renewcommand{\arraystretch}{1.3}
  \begin{small}
  \begin{tabular}{lcccc}
  \toprule
  & \multicolumn{2}{c}{$q\bar{q}\to Zg$} & \multicolumn{2}{c}{$q\bar{q}\to Zgg$} \\
  \cmidrule(lr){2-3}\cmidrule(lr){4-5}
  Parameter & QR & Het & QR & Het \\
  \midrule
  Activation function & GELU & GELU & GELU & GELU \\
  Number of hidden layers & 5 & 5 & 5 & 5 \\
  Hidden nodes & 128 & 128 & 128 & 128 \\
  Input features & 22 & 22 & 30 & 30 \\
  Batch size & 256 & 256 & 256 & 256 \\
  Max learning rate & $2\times10^{-4}$ & $2\times10^{-4}$ & $2\times10^{-4}$ & $2\times10^{-4}$ \\
  Scheduler & Cosine & Cosine & Cosine & Cosine \\
  Number of epochs & 4000 & 4000 & 4000 & 4000 \\
  Output & 49 quantiles & $\mu,\,\sigma$ & 49 quantiles & $\mu,\,\sigma$ \\
  \bottomrule
  \end{tabular}
  \end{small}
  \caption{Network and training hyperparameters for the quantile regression (QR) and
  heteroscedastic (Het) surrogates for both processes.}
  \label{tab:hyper}
\end{table}

In Tab.~\ref{tab:hyper}, we list the hyperparameters used for training the various surrogates.

\section{Supplementary results}

\subsubsection*{Calibration set size}
\label{sec:N_calib}

\begin{figure}[h!]
    \centering
    \includegraphics[width=0.45\linewidth]{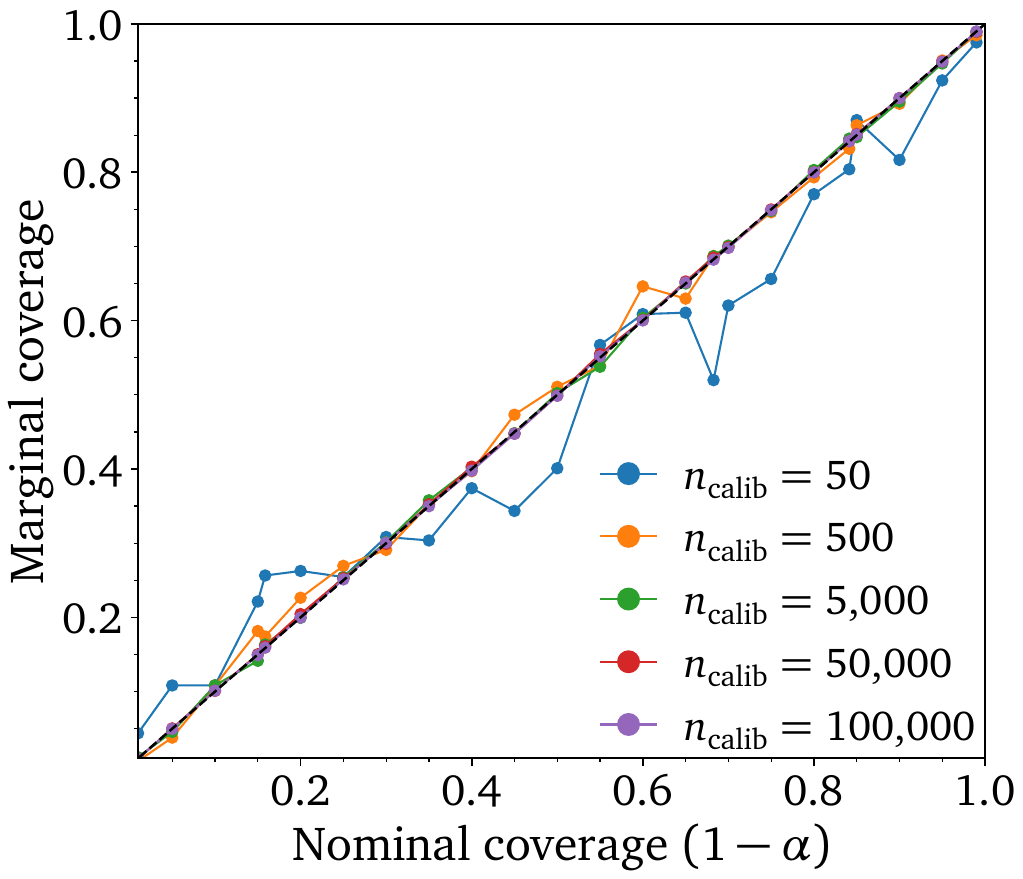}%
    \caption{Marginal coverage vs.\ nominal coverage for FALCON (CQR) for $q\bar{q}\to Zg$,
    with the dashed diagonal representing ideal coverage. The calibration set size
    $n_\text{calib}$ is varied with $n_\text{fresh} = 100$ fixed per probe region.}
    \label{fig:n_calib_marginal_coverage}
\end{figure}

Ref.~\cite{angelopoulos2023} recommends $n_\text{calib}\approx 1000$ as a practical lower
bound, since the coverage uncertainty shrinks as $n^{-1/2}$ and fluctuates too widely at
small $n$.
Fig.~\ref{fig:n_calib_marginal_coverage} confirms this for FALCON: coverage fluctuates
strongly at $n_\text{calib}=50$, converges to the diagonal starting from $n_\text{calib}=500$, and is
indistinguishable from it for $n_\text{calib}=5000$, which we use throughout.

\clearpage
\subsubsection*{Size of fresh probe samples}
\label{app:N_fresh }

\begin{figure}[h!]
    \centering
    \includegraphics[width=0.45\linewidth]{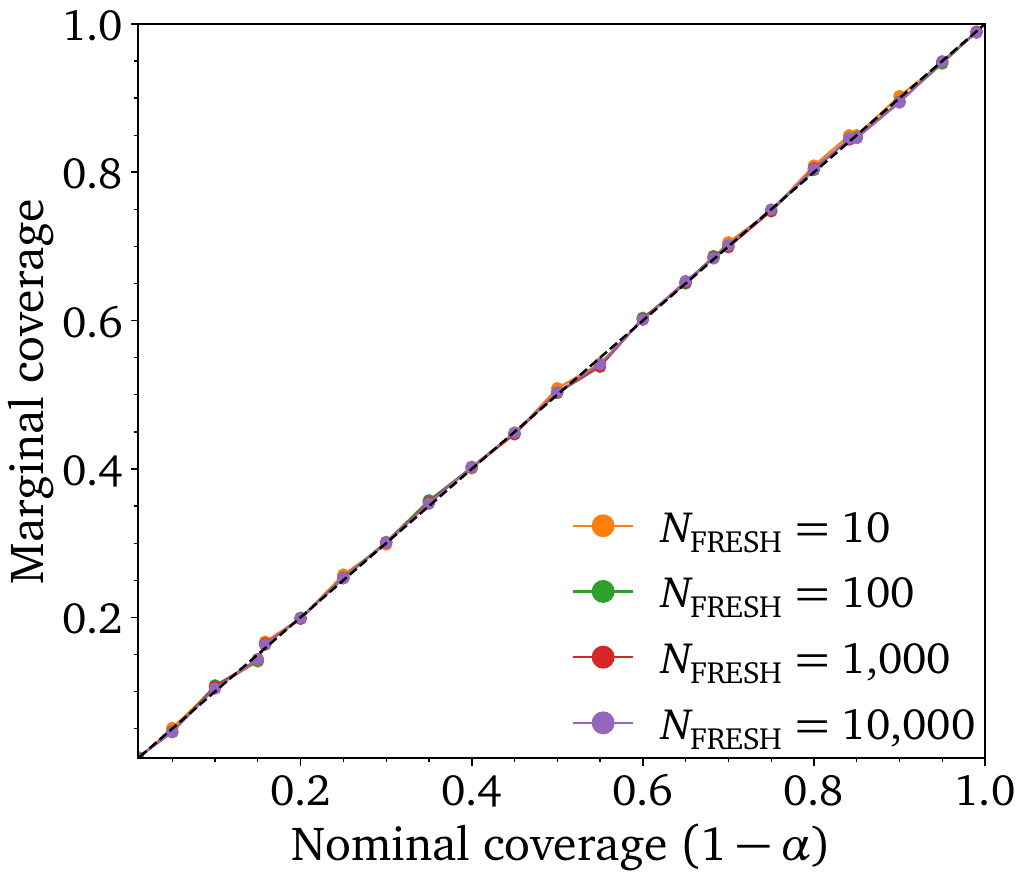}%
    \hfill
    \includegraphics[width=0.45\linewidth]{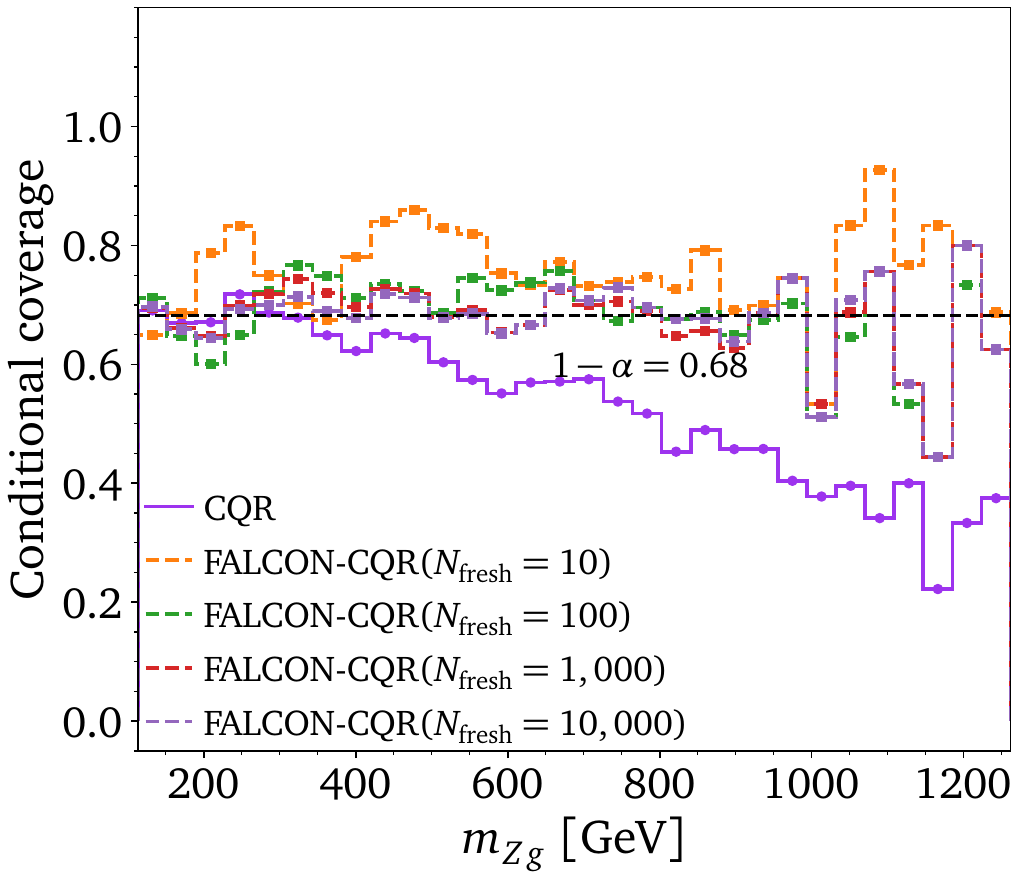}%
    \caption{Coverage of FALCON (CQR) for $q\bar{q}\to Zg$ with varying $n_\text{fresh}$
    with $n_\text{calib} = 5000$ fixed.
    Left: marginal coverage vs.\ nominal coverage, with the dashed diagonal representing exact
    coverage. Right: Conditional 68\% CL coverage as a function of $m_{Zg}$.}
    \label{fig:n_fresh_coverage}
\end{figure}

The marginal guarantee holds for as few as $n_\text{fresh} = 10$ fresh probe events.
Conditional coverage fluctuates at $n_\text{fresh} = 10$ and stabilises near nominal for
$n_\text{fresh} \gtrsim 100$, with no gain from further increases
(Fig.~\ref{fig:n_fresh_coverage}). We use $n_\text{fresh} = 100$ throughout.

\subsubsection*{Coverage in $\cos\theta$}
\label{app:Coverage_in_costheta }

\begin{figure}[h!]
    \centering
    \includegraphics[width=0.45\linewidth]{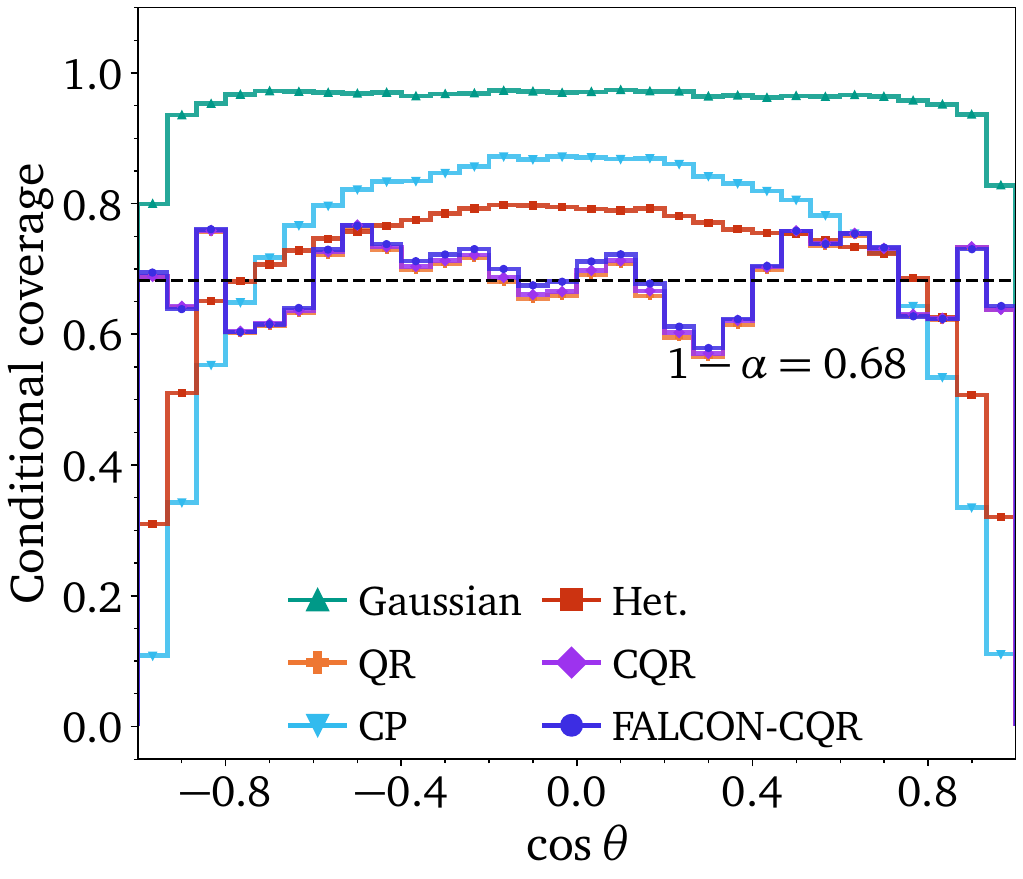}%
    \caption{Conditional 68\% CL coverage for $q\bar{q}\to Zg$ as a function of $\cos\theta$. FALCON is calibrated at 10 evenly spaced probe windows along $m_{Zg}$.}
    \label{fig:zg_invm_interpolation_costheta}
\end{figure}

For the $q\bar{q}\to Zg$ case, we choose probe regions as bins in $m_{Zg}$ (Fig.~\ref{fig:2d_local_coverage}). In Fig.~\ref{fig:zg_invm_interpolation_costheta}, we verify that this leaves the $\cos\theta$ coverage intact. FALCON closely follows CQR and quantile regression, all three maintaining near-nominal coverage across the full angular range. The reason is 
that the amplitude varies mildly with $\cos\theta$, so the CQR quantiles already absorb the angular dependence, leaving no local failure for FALCON to correct. The Gaussian method is underconfident everywhere, whereas the heteroscedastic residual and split methods are underconfident in the bulk and overconfident at the sparse collinear edges.


\subsubsection*{2D phase-space interpolation}
\label{app:2D_interpolation}

\begin{figure}[h!]
    \centering
    \includegraphics[width=0.55\textwidth]{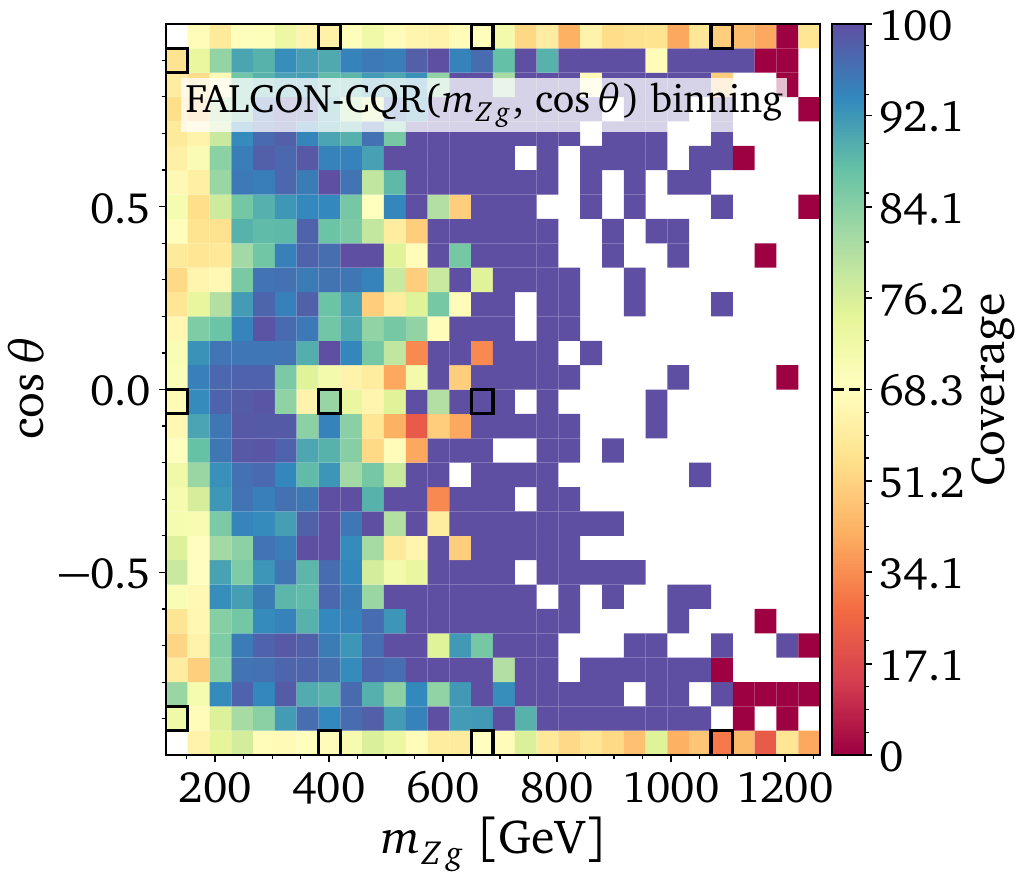}%
    \caption{FALCON with 2D probe interpolation for $q\bar{q}\to Zg$ at the $1\sigma$ level
    ($1-\alpha\approx 0.68$). The phase-space coverage across the $(m_{Zg},\cos\theta)$ plane is shown. The 11 probe regions are indicated by the black boxes.}
    \label{fig:falcon_2d}
\end{figure}

As an alternative to selecting $m_{Zg}$ bins as probe regions for FALCON-CQR --- as shown in Fig.~\ref{fig:2d_local_coverage} ---, we show in Fig.~\ref{fig:falcon_2d} an alternative scheme in which we select small regions in the $(m_{Zg},\cos\theta)$ plane as probe regions. In comparison to Fig.~\ref{fig:2d_local_coverage}, the confidence intervals are underconfident in larger parts of the phase space.

\begin{figure}[h!]
    \centering
    \includegraphics[width=0.44\textwidth]{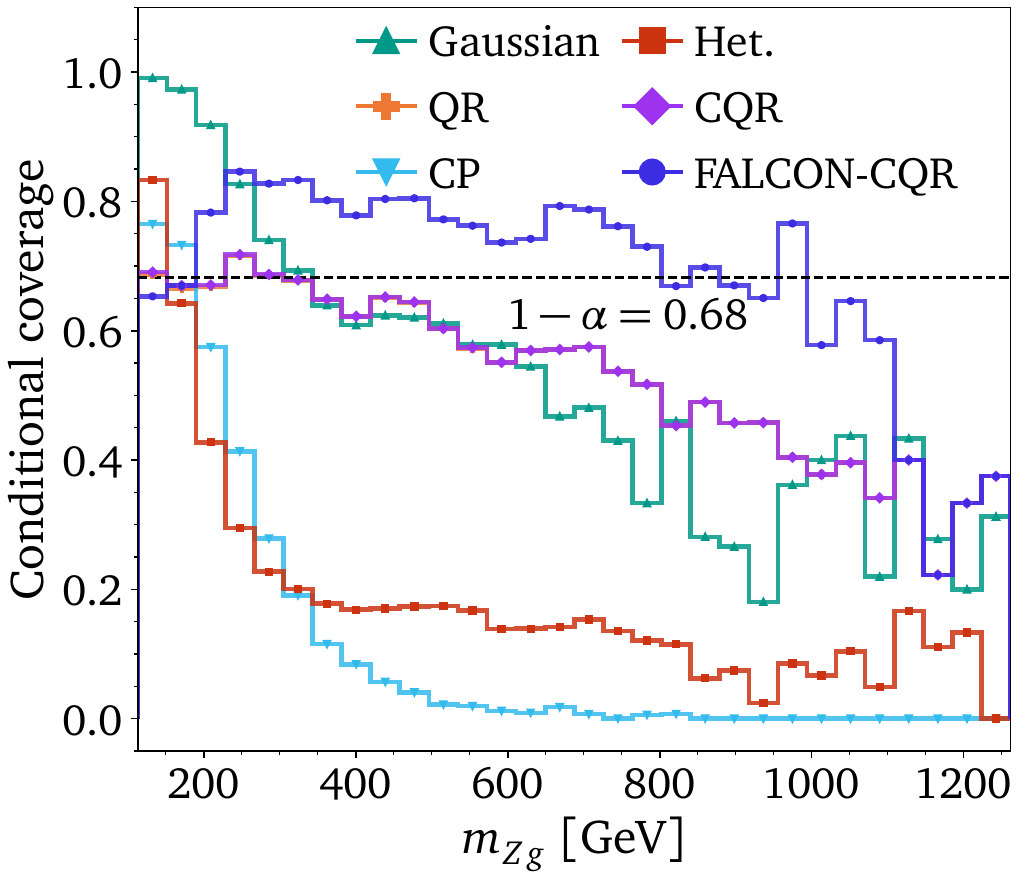}%
    \hfill
    \includegraphics[width=0.44\textwidth]{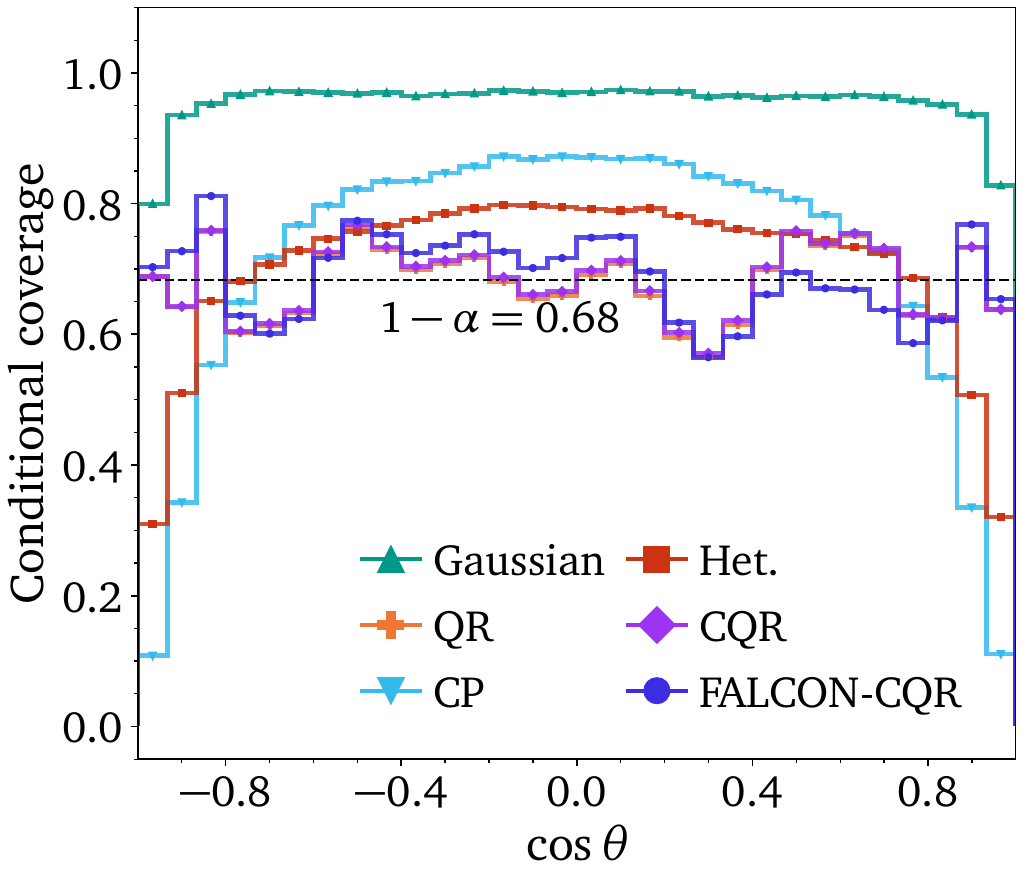}%
    \caption{Conditional 68\% CL coverage from FALCON-CQR with 2D probe interpolation for $q\bar{q}\to Zg$ . Left: as a function of $m_{Zg}$. Right: as a function of $\cos\theta$.}
    \label{fig:falcon_2d_conditional}
\end{figure}

This observation is further supported by the one-dimensional conditional coverage distributions shown in Fig.~\ref{fig:falcon_2d_conditional}. We attribute the inferior performance of the 2D probe-region interpolation compared to the 1D setup to the fact that the underlying quantile regression model already achieves good coverage along the $\cos\theta$ direction. Including this direction in the FALCON interpolation therefore provides little additional benefit and instead introduces interpolation-induced fluctuations, leading to a degradation in performance. The drop in FALCON conditional coverage above $m_{Zg} \gtrsim 1100$~GeV is expected as no probe region is placed beyond this mass, so the tail receives no local correction.



\subsubsection*{$q\bar{q} \to Zg$ Gaussian threshold}
\label{app:gpeak}

As a further test, we apply the conformal methods to the peaked threshold smearing scenario of~\cite{Bahl:2025xvx}. Unlike the globally rescaled noise studied so far, this scenario probes a sharply localized failure mode: the training amplitudes are smeared according to
\begin{align}
  A_\text{train}\sim\mathcal{N}\!\left(A_\text{true},\;
    \varepsilon\,\frac{m_\text{thr}}{\,|m_{Zg}-m_\text{thr}|\,}\,A_\text{true}\right)\,,
  \label{eq:gpeak_noise}
\end{align}
where we set $m_\text{thr} = 200\,\text{GeV}$. The relative noise $\sigma_\text{noise}/A_\text{true} = \varepsilon\,m_\text{thr}/|m_{Zg}-m_\text{thr}|$ diverges at the threshold $m_\text{thr}$ and vanishes in the bulk, so the parameter $\varepsilon$ controls how narrow and how pronounced the noise spike is. We train independent surrogates for $\varepsilon \in \{10^{-3},\,10^{-4},\,10^{-5}\}$ and evaluate conditional and marginal coverage across 30 bins in $m_{Zg}$.

\begin{figure}[t!]
      \centering
      \includegraphics[width=0.3\textwidth]{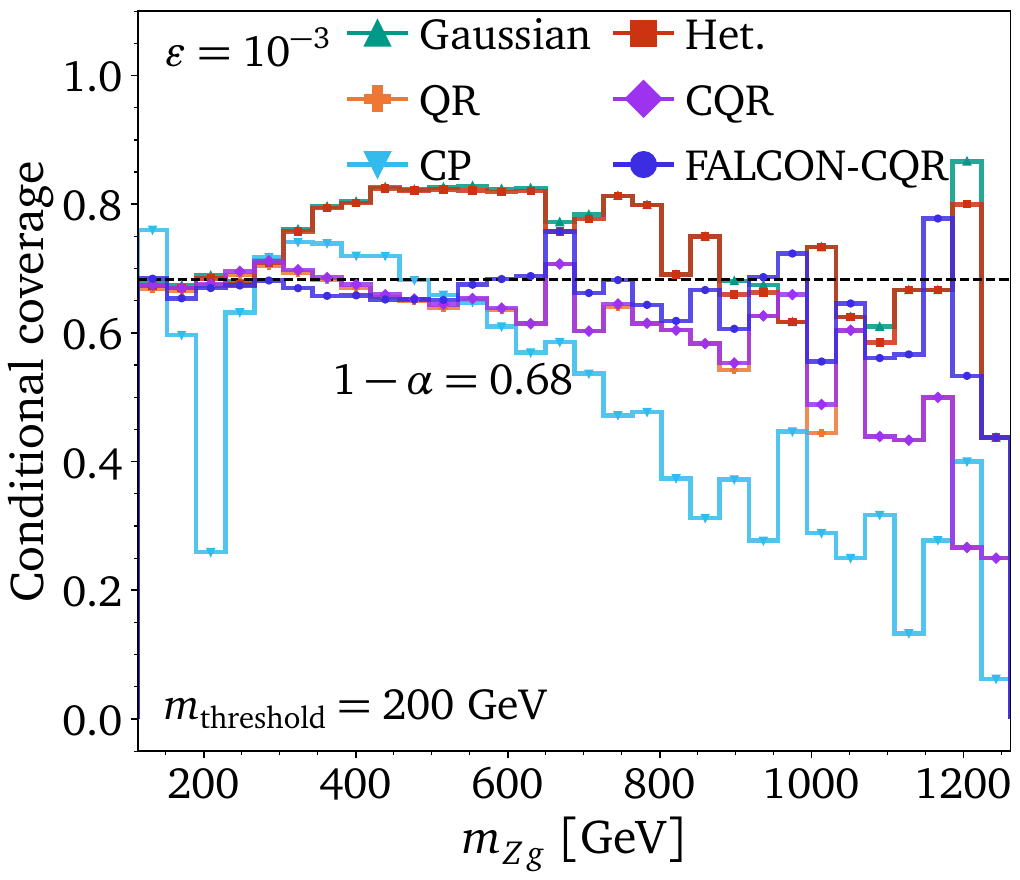}%
      \hfill
      \includegraphics[width=0.3\textwidth]{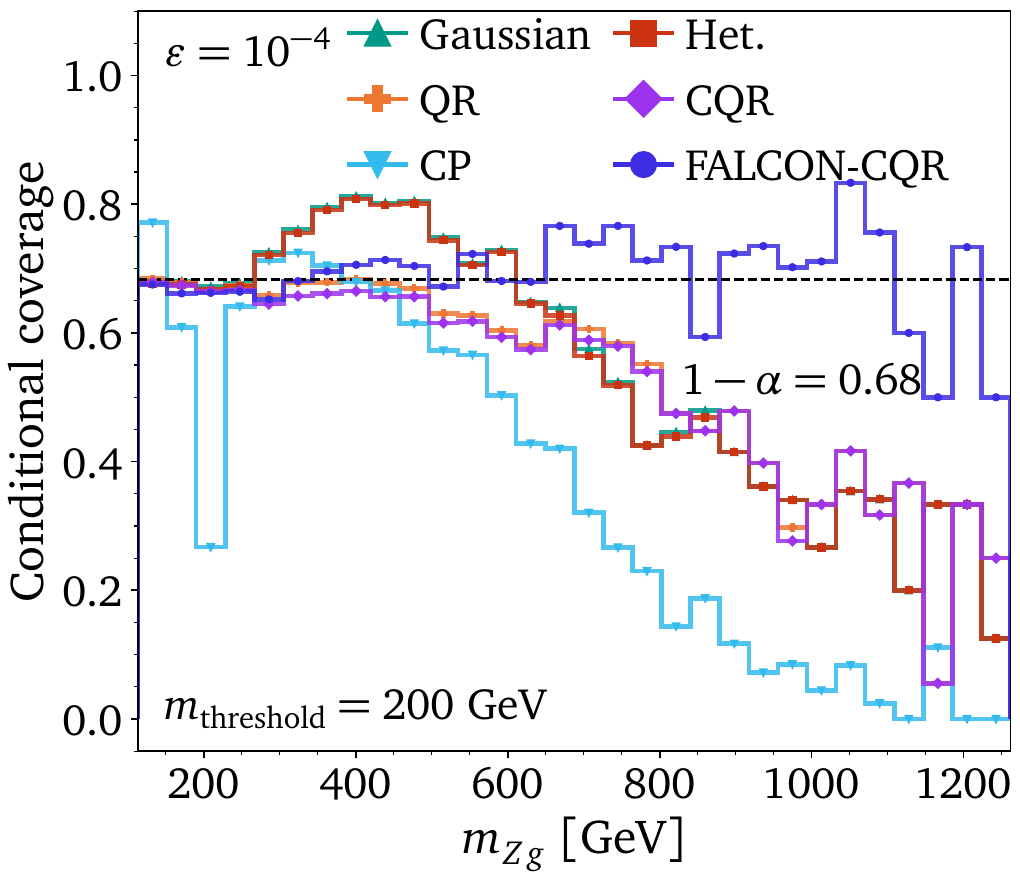}%
      \hfill
      \includegraphics[width=0.3\textwidth]{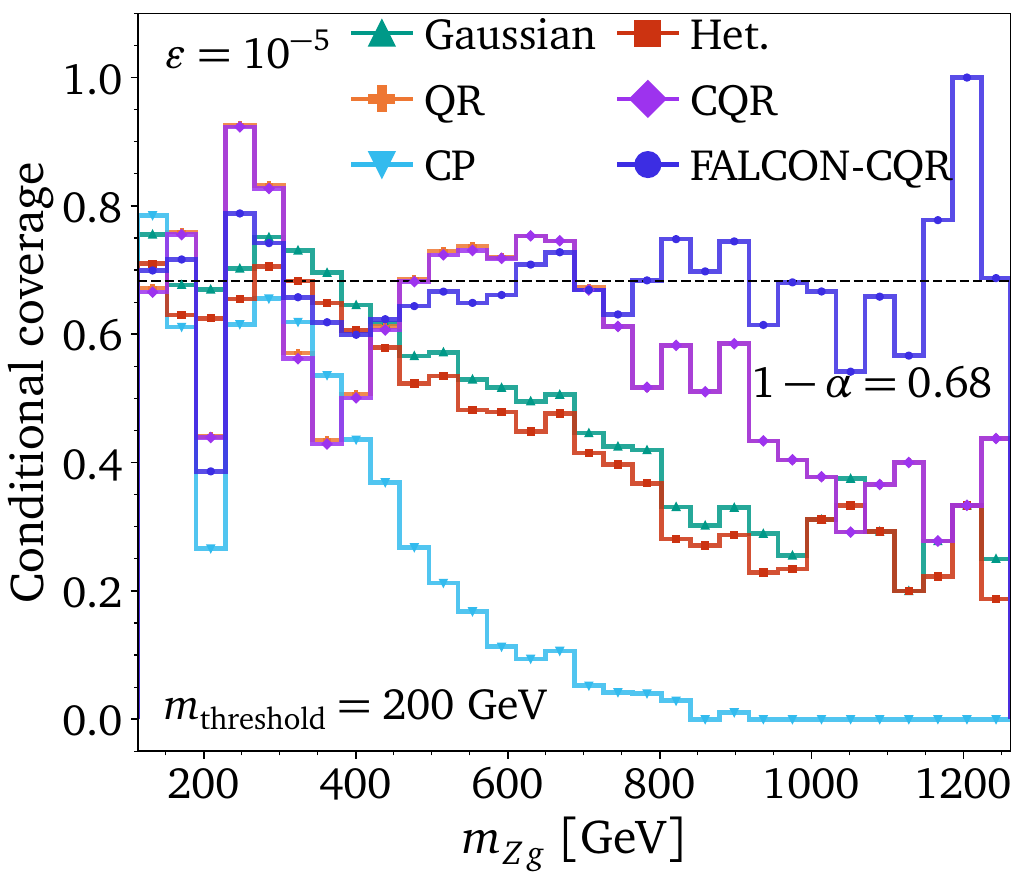}%
      \caption{Conditional 68\% CL coverage for $q\bar{q}\to Zg$ as a function of $m_{Zg}$, for $\varepsilon = 10^{-3}$ (left), $\varepsilon = 10^{-4}$ (center),  and $\varepsilon = 10^{-5}$(right) applied at $m_\text{thr} = 200\,\text{GeV}$ according to Eq.~\eqref{eq:gpeak_noise}.}
      \label{fig:falcon_noise_threshold}
\end{figure}

\begin{figure}[t!]
      \centering
      \includegraphics[width=0.3\textwidth]{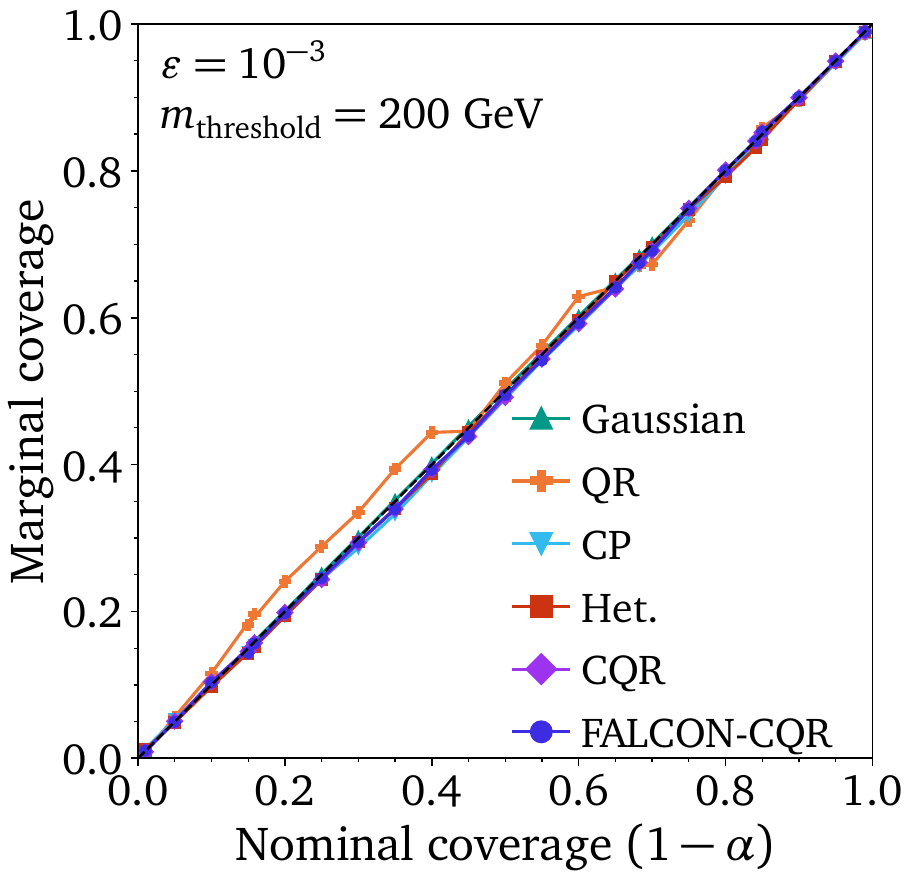}%
      \hfill
      \includegraphics[width=0.3\textwidth]{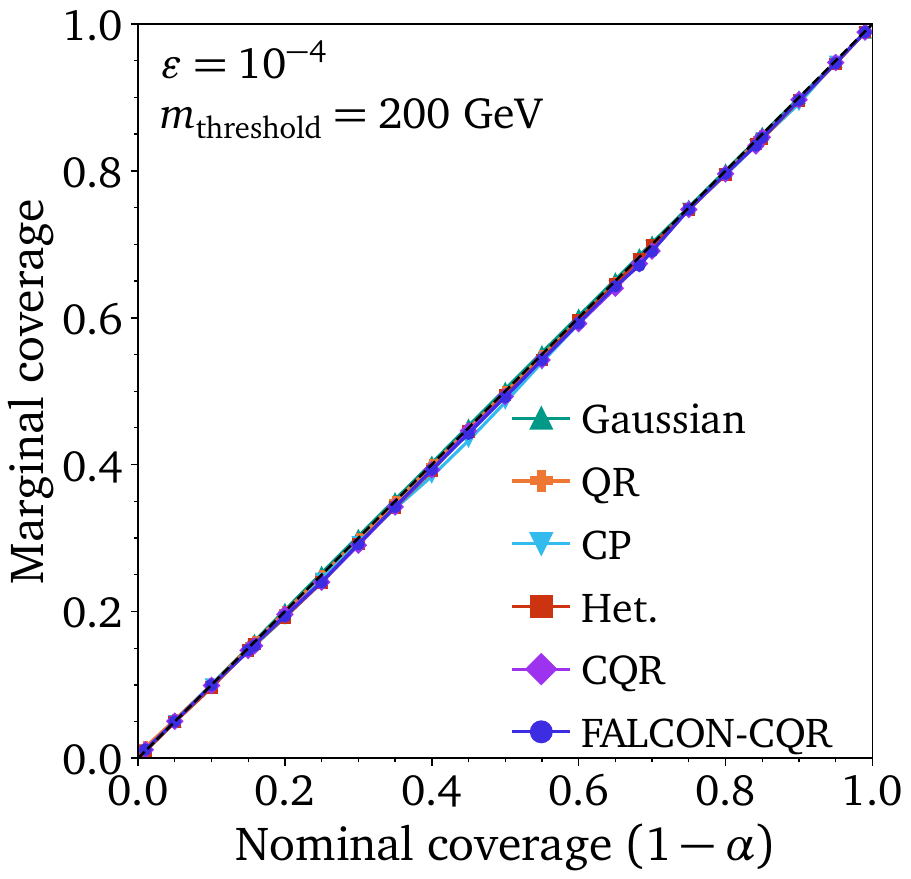}%
      \hfill 
      \includegraphics[width=0.3\textwidth]{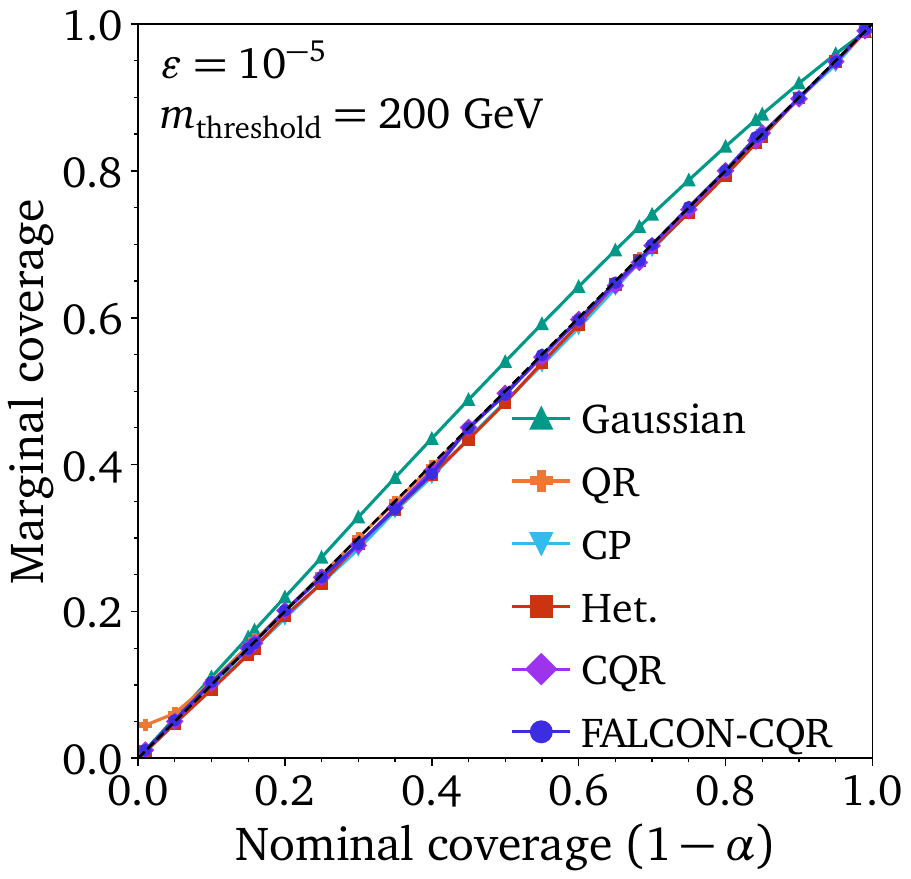}%
      \caption{Marginal coverage vs nominal coverage for $q\bar{q}\to Zg$ with peaked
      threshold noise following Eq.\eqref{eq:gpeak_noise}, for $\varepsilon = 10^{-3}$ (left), $\varepsilon = 10^{-4}$ (center),  and $\varepsilon = 10^{-5}$(right). The dashed diagonal represents exact marginal coverage.}
      \label{fig:falcon_noise_marginal}
\end{figure}

The conditional coverage in Fig.~\ref{fig:falcon_noise_threshold} separates the methods into three groups, according to how they respond to the localized noise spike. Quantile regression and CQR learn the confidence interval $[q_{0.16}(x),\,q_{0.84}(x)]$ directly, without first learning a Gaussian width $\sigma(x)$. Away from the threshold, the noise is negligible --- $\sigma_\text{noise}/A_\text{true}\approx 2\times10^{-3}$ at $300\,\text{GeV}$ --- so both achieve near-nominal conditional coverage at all $\varepsilon$.

The heteroscedastic surrogate instead absorbs the threshold noise into $\sigma$. This enlargement is also propagated towards large $m_{Zg}$, leading to underconfidence between $300$ and $700\,\text{GeV}$. The Gaussian and scaled conformal methods inherit this behavior.

Split conformal prediction on the heteroscedastic surrogate exhibits a distinct failure mode. Near $m_\text{thr}$, the large $\sigma(x)$ suppresses the gradient contributions during training and biases $A_\text{NN}$ at the threshold. The global $q_\alpha$ is dominated by bulk events and cannot compensate for this local bias. As a consequence, the calibration interval undercovers at $m_\text{thr}$, reaching only ${\approx}0.27$ at all $\varepsilon$. Above $600\,\text{GeV}$ the statistics are too sparse, and the coverage collapses, consistent with the no-noise baseline of Fig.~\ref{fig:invm_cqr_conditional_coverage}.

Finally, FALCON places its two innermost probe regions at $m_{Zg}\simeq 132$ and $m_{Zg}\simeq 247\,\text{GeV}$. At $\varepsilon = 10^{-3}$ the spike is wide enough for both adjacent probe regions to register it, and FALCON maintains $1-\alpha = 0.68$ across the full mass range. At $\varepsilon = 10^{-4}$ and $10^{-5}$, however, the spike becomes narrower than the probe spacing. Consequently, neither probe captures the local failure mode. The interpolated correction is then insufficient at $m_\text{thr}$ and the coverage dips to ${\approx}0.4$. This illustrates the expected resolution limit of the interpolation. FALCON corrects local features only down to the scale set by the probe-window spacing.

The marginal coverage in Fig.~\ref{fig:falcon_noise_marginal} confirms that all conformal methods satisfy the marginal guarantee at every $\varepsilon$, as required by Eq.~\eqref{eq:cp_sandwich}. The Gaussian method tracks the diagonal at $\varepsilon = 10^{-3}$ and $10^{-4}$, where the noise dominates and the pull $(A_\text{train}-A_\text{NN})/\sigma$ is approximately Gaussian. At $\varepsilon = 10^{-5}$ the heteroscedastic model overestimates $\sigma$ near the high-statistics threshold, leading to small underconfidence. Quantile regression deviates slightly above the diagonal only at $\varepsilon = 10^{-3}$, where the heavy tails near the threshold widen the learned quantile intervals.



\subsubsection*{Pull distribution}
\label{app:Pull_distribution}

\begin{figure}[h!]
    \centering
    \includegraphics[width=0.45\linewidth]{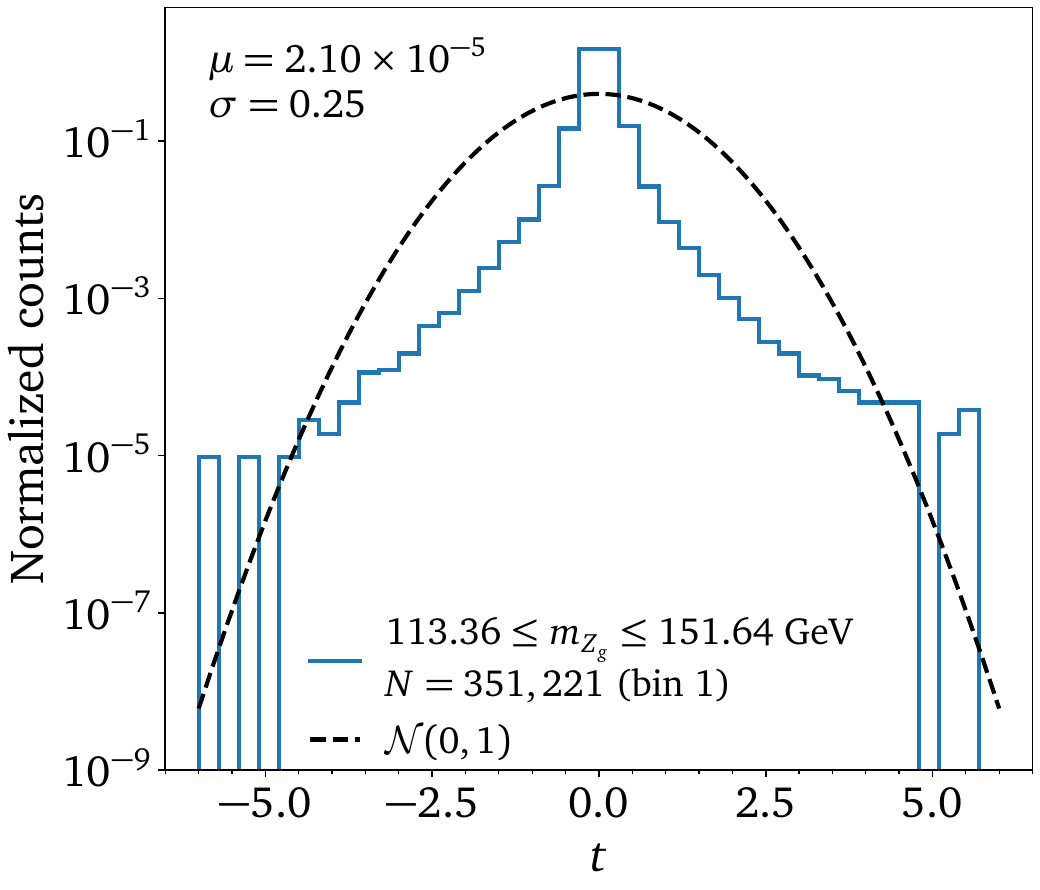}%
    \hfill
    \includegraphics[width=0.45\linewidth]{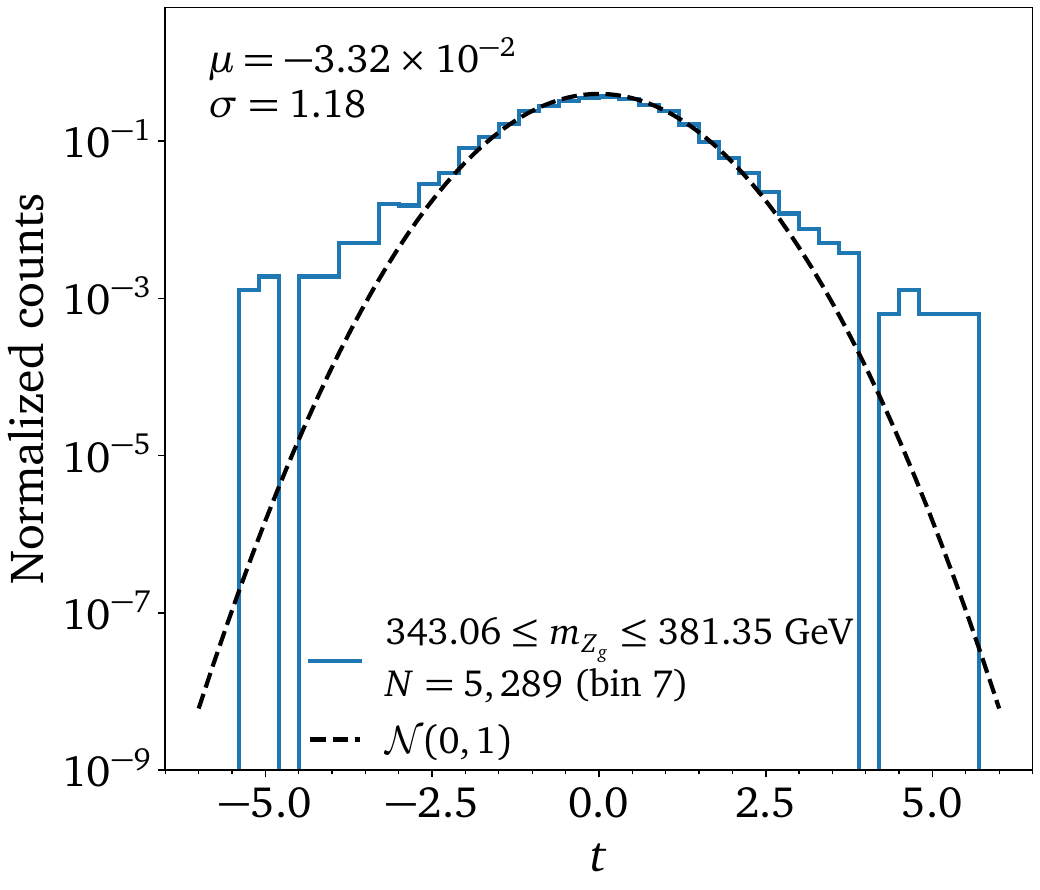}%
    \caption{Pull distributions for two $m_{Zg}$ slices based on the $q\bar{q}\to Zg$ heteroscedastic surrogate.}
    \label{fig:pull_distribution}
\end{figure}

To further illustrate the Gaussian coverage observations of Figs.~\ref{fig:invm_cqr_conditional_coverage} and~\ref{fig:2d_local_coverage}, we investigate the pull
\begin{align}
    t(x) = \frac{A_\text{NN}(x) - A_\text{true}(x)}{\sigma(x)}\;.
    \label{eq:pull}
\end{align}
If the residuals are Gaussian distributed and $\sigma$ is correctly calibrated, the pull distribution will follow a unit Gaussian. We investigate it in two $m_{Zg}$ slices. For $m_{Zg} \approx 113~...~152$~GeV, the pull plot is non-Gaussian despite sufficient statistics, confirming that the underconfidence of the heteroscedastic surrogate in this region is driven by a breakdown of the Gaussian assumption. For $m_{Zg} \approx 343~...~ 381$~GeV, the pull plot follows a standard normal distribution in the bulk region, consistent with the near-nominal coverage seen in both the 1D and phase-space coverage plots across this mass range.


\clearpage
\bibliography{tilman,refs}

\end{document}